\documentclass[aps,prd,twocolumn,reprint,floatfix,superscriptaddress]{revtex4-2}
\usepackage{tabularx}
\usepackage{graphicx}
\usepackage{amsmath}
\usepackage{amssymb}
\usepackage{comment}
\usepackage{xcolor}
\usepackage{xspace}
\usepackage{multirow}
\usepackage{natbib}
\usepackage{hyperref}
\usepackage{ulem}
\usepackage{dcolumn}
\usepackage{siunitx}
\usepackage{makecell}
\usepackage{subcaption}
\usepackage{cleveref}
\usepackage{orcidlink}

\newcommand{\algoname}{\texttt{ABNORMAL}\xspace}

\newcommand{\nblks}{N_{\rm blks}}
\begin{document}
\title{Using normal to find abnormal: AI-based anomaly detection in gravitational wave data}
\author{Yi-Yang Guo,\orcidlink{0009-0003-0678-2488}}
\email{yguo173@ucsc.edu}
\affiliation{
    Lanzhou Center for Theoretical Physics, Key Laboratory of Theoretical Physics of Gansu Province, Key Laboratory for Quantum Theory and Applications of the Ministry of Education, Lanzhou University, Lanzhou, Gansu 730000, China
}
\altaffiliation{Department of Electrical and Computer Engineering, University of California, Santa Cruz, CA 95064}
\affiliation{
    Institute of Theoretical Physics $\&$ Research Center of Gravitation, School of Physics Science and Technology, Lanzhou University, Lanzhou 730000, China
}
\affiliation{
    Gansu Provincial Research Center for Basic Disciplines of Quantum Physics, Lanzhou University, Lanzhou, Gansu 730000, China
}
\author{Soumya D.~Mohanty,\orcidlink{0000-0002-4651-6438}}
\email{soumya.mohanty@utrgv.edu}
\affiliation{Department of Physics and Astronomy, The University of Texas Rio Grande Valley, One West University Blvd.,
Brownsville, Texas 78520}
\affiliation{Department of Physics, IIT Hyderabad, Kandi, Telangana-502284, India}
\author{Xie Qunying,\orcidlink{****}}
\affiliation{Schoool of Information Science $\&$ Engineering, Lanzhou University, Lanzhou}
\email{xieqy@lzu.edu.cn}
\author{Yu-Xiao Liu,\orcidlink{****}}
\affiliation{
    Lanzhou Center for Theoretical Physics, Key Laboratory of Theoretical Physics of Gansu Province, Key Laboratory for Quantum Theory and Applications of the Ministry of Education, Lanzhou University, Lanzhou, Gansu 730000, China
}
\affiliation{
    Institute of Theoretical Physics $\&$ Research Center of Gravitation, School of Physics Science and Technology, Lanzhou University, Lanzhou 730000, China
}
\affiliation{
    Gansu Provincial Research Center for Basic Disciplines of Quantum Physics, Lanzhou University, Lanzhou, Gansu 730000, China
}

\email{****}
\begin{abstract}
    The detection and classification of anomalies in gravitational wave data plays a critical role in improving the sensitivity of searches for signals of astrophysical origins. We present \algoname (AI Based Nonstationarity Observer for Resectioning and Marking AnomaLies), a deep neural network (DNN) model for anomaly detection that is trained exclusively on simulated Gaussian noise. By removing dependence on real data for training, the method resolves a circular paradox in anomaly detection: training on real data implicitly involves prior segregation of stationary from non-stationary data but this is not possible unless all anomalies are detected first. \algoname\ is an autoencoder-based DNN, commonly used in anomaly detection, with the key innovation that it is trained to predict statistical features of noise rather than reconstructing the noise time series themselves. The statistical features are obtained by applying Gabor and Wavelet filter banks, which implement time-frequency analysis, and are subsequently combined through multi-view fusion using a dual-path architecture. 
    We quantify the performance of our method on simulated and real LIGO data.
    Application to data from the O1 to O3b observational runs uncovers a rich landscape of anomalies over different timescales, including many that do not fit within known classes.

    \end{abstract}
\maketitle


\section{Introduction}
\label{sec:intro}

Gravitational Wave (GW) astronomy has achieved spectacular success since the direct detection of the first signal, GW150914~\cite{PhysRevLett.116.061102}, from a binary black hole (BBH) merger in $2015$. Across the the observing runs O1, O2, O3a, and O3b carried out so far~\cite{abbott2021gwtc}, with O4 currently in progress, the network of the twin LIGO detectors~\cite{2016PhRvL.116m1103A} in U.S.A and Virgo~\cite{2013ASPC..467..151D} in Europe have detected $182$ merger events. The majority of the detected events so far, collectively called compact binary coalescences (CBCs), are BBH mergers but there are a handful that involve neutron stars.  With the Japanese KAGRA~\cite{2012CQGra..29l4007S} detector joining observations starting in late O3 and the construction of LIGO-India~\cite{2013IJMPD..2241010U} underway, the build-out of the worldwide network of $2^{\rm nd}$ generation detectors is nearing completion. This will further increase the overall search sensitivity and rate of GW discoveries along with significantly better parameter estimation accuracies.

GW data is noise dominated and, hence, sophisticated signal processing and statistical data analysis methods are needed to search for weak and infrequent signals of astrophysical origin. While the mathematical form of data analysis methods, such as matched filtering\cite{owen1999matched,schutz1994searching,dhurandhar2004data,sathyaprakash11997signal}, used in the searches for GW signals are derived from idealized assumptions of stationarity and Gaussianity of the noise, real data frequently deviates from these assumptions. In this paper, we refer to all deviations from stationarity as anomalies — a broad term that may include GW signals when there is no risk of confusion. 

Anomalies arising from environmental or instrumental sources, occur on a large range of timescales and in various forms. Short duration ($\lesssim 1$~sec) ones, commonly called glitches, cause a degradation in the sensitivity of searches for transient GW signals~\cite{Abbott_2018}. In some cases, glitches can also mask GW signals, as in the case of the first observed double neutron star (DNS) merger  GW170817~\cite{PhysRevLett.119.161101}. Longer duration anomalies include, but are not limited to, narrowband noise features that drift in frequency (wandering lines), non-GW chirps, and broadband excess noise. Some anomalies can be correlated with data from auxiliary sensors and this is used to produce data quality
flags~\cite{slutsky2010methods} that guide the selection of data in search pipelines.  

The LIGO-Virgo-KAGRA collaboration uses several established GW search pipelines. Among these are Coherent WaveBurst (CWB)~\cite{klimenko:2005,PhysRevD.93.042004}, which is primarily used for GW burst detection, while pipelines such as \texttt{PyCBC}~\cite{biwer2019pycbc}, GSTLal~\cite{cannon2021gstlal}, MBTA~\cite{adams2016low}, and SPIIR~\cite{hooper2012summed} use matched filtering for CBC detection. 
In these pipelines, the use of multiple detectors plays a crucial role in distinguishing GW signals from non-GW anomalies. Beyond GW signals, the Omicron~\cite{robinet2007omicron} pipeline is used for glitch detection and provides the information underlying data quality flags~\cite{rodriguez2007reducing} that guide all GW searches. Several methods are used for follow ups of the candidates produced by the search pipelines. For example, Bayeswave~\cite{cornish2015bayeswave} leverages data from multiple detectors to differentiate between glitches and genuine GW signals. To assist experimentalists in identifying hardware issues~\cite{2010NIMPA.624..223A}, it is essential to classify the non-GW anomalies. For this task, the citizen science project Gravity Spy~\cite{zevin2017gravity} has identified $23$ distinct classes of glitches, highlighting the diverse nature of anomalies in GW data.

The pipelines above rely on the theoretically problematic assumption that some portion of the data being analyzed is stationary. This portion is then typically used for the training of pipeline components, such as whitening filters~\cite{cuoco2001line,cuoco2004whitening,tsukada2018application} based on the estimated Power Spectral Density (PSD) of noise. However, given data that has anomalies, such a portion of the data cannot be found unless the anomalies in the data are detected first, leading to a circular conundrum. In terms of the commonly used terminology in the literature on anomaly detection~\cite{chandola2009anomaly}, the problem described above is that of finding normal portions of data, a task called input data labeling,  from a mix of normal and abnormal data.

Practical implementations attempt to sidestep the input data labeling problem by using outlier-resistant statistical measures—like the median instead of the mean—to estimate training information such as power spectral densities (PSDs) \cite{usman2016pycbc}. However, this merely shifts the burden to assumptions about other anomaly properties, including their timescales. There exist alternative classical statistical approaches for anomaly detection, such as non-parametric change point detection methods like KSCD~\cite{mohanty2005progress} and BlockNormal~\cite{matsakis2010active}, that attempt to minimize prior assumptions about stationarity but they require careful tuning of algorithmic parameters, such as time-frequency resolution or block length. Moreover, the theoretically infinite variety of anomalies, with new types continuously emerging, makes it difficult to tune and adapt classical methods. 


 In recent times, Deep Learning using artificial neural networks
 has emerged as a highly successful paradigm across various domains~\cite{tan2018artificial,dargan2020survey,goyal2022inductive,deng2018artificial,sallab2017deep,jaderberg2019human} for tasks where the data characteristics are too complex to be captured by simple assumptions and models. Deep learning models encompass several specialized architectures~\cite{baraniuk2020science,liu2017survey,aggarwal2018neural}, each designed for a specific type of analysis: Convolutional Neural Networks (CNNs) excel at image processing through hierarchical feature extraction, Recurrent Neural Networks (RNNs)\cite{medsker2001recurrent} and Transformers\cite{vaswani2017attention} handle sequential data by capturing temporal dependencies, while autoencoders\cite{rumelhart1986learning} learn compressed representations of data, making them particularly suitable for anomaly detection through their ability to reconstruct normal patterns. In GW data analysis, autoencoder-based approaches have shown promise by learning to characterize normal detector behavior during training, subsequently identifying anomalies through elevated reconstruction loss. Recent work in this direction includes GWAK~\cite{Raikman2023GWAKGA}, which employs recurrent autoencoders for anomaly detection in data from a detector network, expanding upon earlier efforts~\cite{Morawski2021AnomalyDI,Moreno2021SourceagnosticGD} that established the viability of deep learning for GW data characterization.

While autoencoder-based anomaly detection approaches show promise, existing implementations have the following issues that potentially reduce their generalization capacity. One is the conceptual challenge associated with input data labeling, as previously discussed, and the other is the reliance on labeled anomaly classes in the training data to improve performance. A downside of the latter is that the method becomes dependent on preidentified glitch classes, but these classes and their prevalence in the data change over time~\cite{wu2024advancingglitchclassificationgravity, ferreira2024using}, requiring repeated re-tuning of the method. The labeling of the glitches may also have errors that propagate through the model training and adversely affect inference. In addition, a technical limitation common to existing autoencoder-based approaches is that they are trained to directly predict a noise time series. This approach is problematic as it attempts to predict specific realizations of an intrinsically unpredictable random process. 
Similarly, training an autoencoder-based model with real detector data alongside simulated noise can potentially introduce biases from the presence of undetected anomalies in the training dataset.

To address these limitations, We present \algoname (AI Based Nonstationarity Observer for Resectioning and Marking AnomaLies), a novel approach to GW anomaly detection. While building on the autoencoder framework, our method introduces two key innovations: first, we train exclusively on simulated stationary noise from an ensemble of PSDs, eliminating the input data labeling problem and potential biases from real data; second, we focus on predicting statistical summaries of the noise rather than raw time series values, providing more stable and meaningful feature representations. The architecture of the model employs a multi-view fusion strategy, combining features extracted from different sets of bandpass filters, while supporting variable time scales for feature generation to capture anomalies across different temporal ranges. Formally, \algoname belongs to the class of semi-supervised anomaly detection methods that are trained only on normal data~\cite{chandola2009anomaly}. We go a step further by confining the training exclusively to simulated data. 

Our results show that it is indeed feasible to train a model exclusively on simulated stationary noise and use it to detect anomalies in real data at a useful sensitivity. We believe that future enhancements, such as better engineering of input features and better network architectures, could significantly enhance the performance of this approach. The remainder of this paper is organized as follows. Section~\ref{sec:method} describes the data model used in the paper and the generation of features that comprise the input to \algoname. Section~\ref{sec:datasets} details the generation of simulated training and testing datasets. The architecture and training procedure of our deep neural network model is presented in Section~\ref{sec:dnn_model}. 
The application of \algoname to real data requires considerations, such as the removal of lines and estimation of detection thresholds, that are described in Sec.~\ref{sec:results_threshold}.
Section~\ref{sec:results} presents our main results based on both simulated and real GW data. Finally, we summarize our findings and discuss future directions in Section~\ref{sec:conclusions}.

\section{Data model and Input features}
\label{sec:method}

GW strain data from a single detector can be modeled as $\overline{y}=\overline{s}+\overline{n}$, where $\overline{n}$ and $\overline{s}$ denote the detector noise and a potential anomaly, respectively. In the design of conventional data analysis methods, $\overline{n}$ is assumed to be a realization of a stationary Gaussian random process. Within our framework, we consider any $\overline{s}$ to be an anomaly, regardless of its origin, and use this term interchangeably with the term signal. It is important to note that non-GW anomalies can also arise from deviations of $\overline{n}$ from stationarity that cannot be decomposed into an additive signal model. For example, non-linear couplings in the detector~\cite{abbott2016characterization} can create harmonically spaced disturbances in the noise spectrum. Temporal variations in these couplings produce non-stationarities that violate the additive signal model as they cannot be separated into a distinct $\overline{s}$ component while maintaining the stationarity of $\overline{n}$.
Although the additive model above is not assumed for training the model in \algoname, since it is trained purely on noise, we will confine its testing in this paper to the additive model for simplicity.

Autoencoders for anomaly detection are typically trained to reconstruct non-anomalous data that comes in a finite number of forms. However, non-anomalous GW data are realizations of a noise process and, as such, are infinitely diverse. Therefore, using them directly in training an autoencoder has a significant negative impact on its generalization ability. We circumvent this problem by using statistical features, described below, that show much better stability for stationary noise but deviate systematically during anomalous events.

\subsection{Filter banks}
\label{sec:filter_banks}
In the class of anomalies under the additive model, the dominant observed population has the energy in $\overline{s}$ localized in both time and frequency. Therefore, the majority of non-parametric methods used in the LVK for detecting such signals rely on some variant of time-frequency analysis such as the spectrogram (or the Gabor transform~\cite{gabor1946electrical}), wavelet~\cite{daubechies1988orthonormal}, or the constant Q-transforms~\cite{brown1991calculation, chatterji2004multiresolution}. 
Given the usefulness of time-frequency transforms, we adopt the same idea for constructing the input features. Since our input features are generated using time series data, we implement the time-frequency transforms using banks of bandpass filters. This is done using two separate filter banks, corresponding to the Gabor transform and the continuous wavelet transform (CWT).

\begin{figure}[t]
    \centering
    \includegraphics[width=\linewidth]{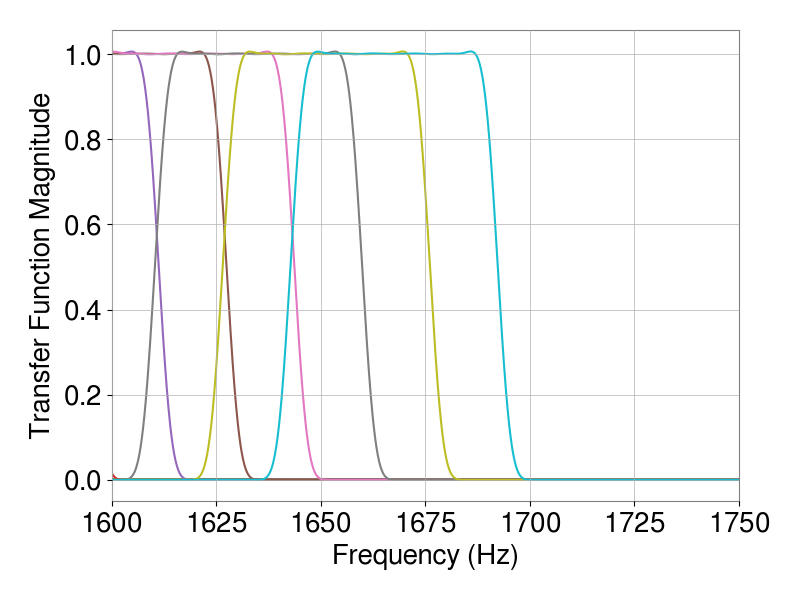}
    \caption{Magnitudes of the transfer functions of the FIR bandpass filters used in the Gabor filter bank (GFB).  For visual clarity, only the filters towards the end of this range, between $1600$~Hz and $1750$~Hz, are shown here.} 
    \label{fig:filter_banks_multi}
\end{figure}

The Gabor transform is equivalent to a set of filters, which we call the Gabor filter bank (GFB), having the same passband widths and spaced uniformly apart in their center frequencies.
 Based on experiments described later in the paper, we use $100$ zero-phase FIR bandpass filters, each of order $1001$, 
 with center frequencies of the passbands spaced uniformly within $[30, 1700]$~Hz. These filters are characterized by a consistent $-3$~dB bandwidth of $47$~Hz, passband ripple of $3.1$~dB, and stopband attenuation exceeding $120$~dB. The passbands of adjacent filters have an overlap of $75\%$. Figure \ref{fig:filter_banks_multi} shows the transfer function magnitudes of the filters, illustrating their uniform bandwidths and shifts in frequency.

\begin{figure}[t]
    \centering
    \includegraphics[width=\linewidth]{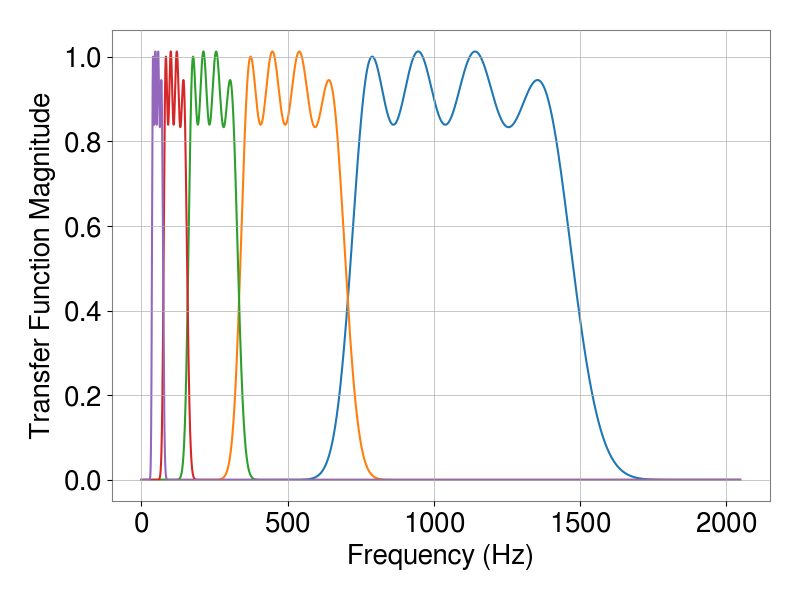}
    \caption{Magnitudes of the transfer functions of the Wavelet Filter bank (WFB) filters based on Morlet wavelets. }
    \label{fig:filter_banks_wavelet}
\end{figure}

For the CWT, we use filters corresponding to Morlet wavelets with $400$ wavelet scales  spaced geometrically from $6.4$ to $224$.
The bandpassed time series are obtained using the inverse CWT (ICWT) through the \texttt{ssqueezepy} library. For this, the wavelet coefficients are grouped into sets of $100$ scales with overlap and, for each set, the ICWT is computed using only the coefficients in the set.
Figure \ref{fig:filter_banks_wavelet} shows the transfer function magnitudes of the 5 resulting bandpass filters that are equivalent to the ICWT reconstruction.  We call this set of filters the wavelet filter bank (WFB). Unlike the uniform bandwidth GFB, the WFB provides natural multi-resolution analysis with bandwidths that vary across the frequency spectrum, offering better time resolution at higher frequencies and better frequency resolution at lower frequencies.

\subsection{Input features}
\label{subsec:features}
For each bandpass filter, a set of input features are generated from a finite length segment of the filtered time series as follows. The time series segment is first normalized by splitting it into $20$ contiguous blocks of equal length, computing the standard deviation of the time series in each block, and scaling the entire segment by the median of this set of standard deviations.  This in situ normalization step, which is independent of any assumption of stationarity, should not be confused with the requirement of whitening in other approaches to anomaly detection -- Normalization of inputs is a general requirement for neural networks due to the training process that depends on gradient descent. Due to the large dynamic range in the PSD of GW detectors, a large part of the normalization has to be performed before feeding them into the network. Our use of the median matches the Power Spectral Density (PSD) estimation method used in CBC searches~\cite{usman2016pycbc} for robustness against glitches. 

After filtering and normalization, we transform each segment into a set of feature vectors as follows. (i) The segment is split into $N_{\text{blks}}$ non-overlapping blocks, with the block size determining the time scale of anomalies to which \algoname is sensitive. (Note that  $N_{\rm blks}$ is a tunable parameter of \algoname while the number of blocks used for normalization is fixed.) For each block, we compute the mean and the variance of the time series.
(ii) A histogram of the full segment is constructed using $N_{\text{blks}}$ bins, capturing the global distribution of the data values in the segment. The histogram counts are normalized using their $L_2$ norm to ensure a uniform amplitude scale across the different frequency bands. Since the number of bins is fixed, excessively large amplitude values in the segment, such as from a glitch, will result in the histogram being compressed towards the center. For the block-wise variances associated with the time series from the WFB, we find it necessary to introduce a constant empirical attenuation factor of $10^{-2}$ for improved performance. 
This factor improves discrimination between genuine anomalies and normal fluctuations: strong glitches still produce clear signals even after attenuation, while the reduced scale helps suppress the naturally large variance fluctuations that can occur in stationary noise. 

The features described above are packed into two matrices for input to the model. 
For the WFB, the matrix has $15$ rows, corresponding to the $3$ statistical features for each of $5$ frequency bands (c.f., Fig.~\ref{fig:filter_banks_wavelet}), and $N_{\rm blks}$ columns. Similarly, for the GFB, we originally have $300$ rows ($100$ bands $\times$ 3 statistical features) and $N_{\rm blks}$ columns. However, it is reshaped to $15$ rows and $20\times N_{\rm blks}$ columns to ensure dimensional compatibility with the WFB features for reasons explained later (see Sec.~\ref{sec:model_arch}).

\begin{figure*}
    \centering
    \includegraphics[width=\textwidth]{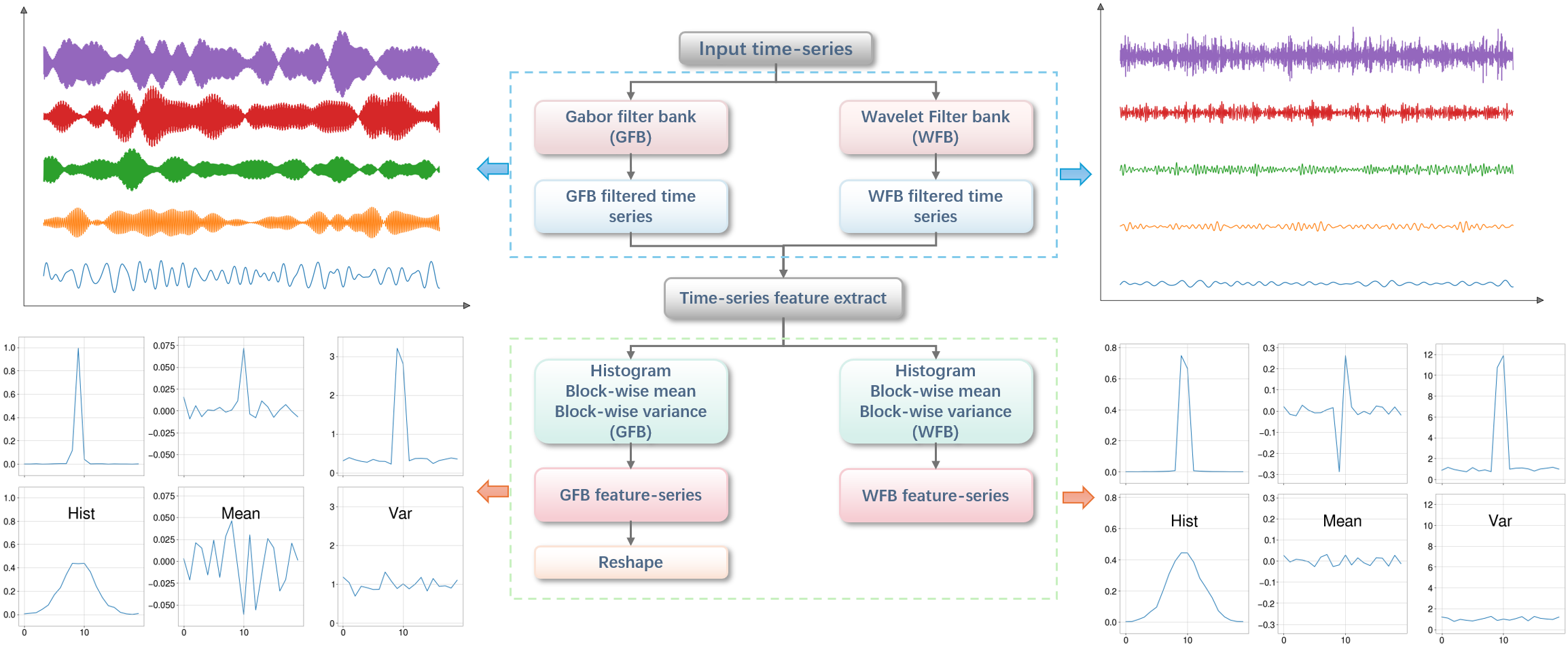}
    \caption{Overview of the feature generation pipeline in \algoname. The process begins with the input time-series (top) which is simultaneously processed through two parallel filter banks: (left) Gabor filter bank (GFB) (illustrated with a GFB containing only $5$ filters) and (right) Wavelet Filter bank (WFB). The resulting filtered time series (middle level) are then normalized and analyzed to extract statistical features (third level): histograms, block-wise means, and block-wise variances. The GFB features undergo a reshaping operation (fourth level) before both sets of features are sent to the \algoname DNN model. The $2$-by-$3$ grid of lower level plots show the extracted features for a single filtered time series containing (top row)  a glitch and (bottom row) only stationary noise, illustrating how an anomaly affects the features.}
    \label{fig:features_plan}
\end{figure*}

\begin{figure*}
    \centering
    \includegraphics[width=\textwidth]{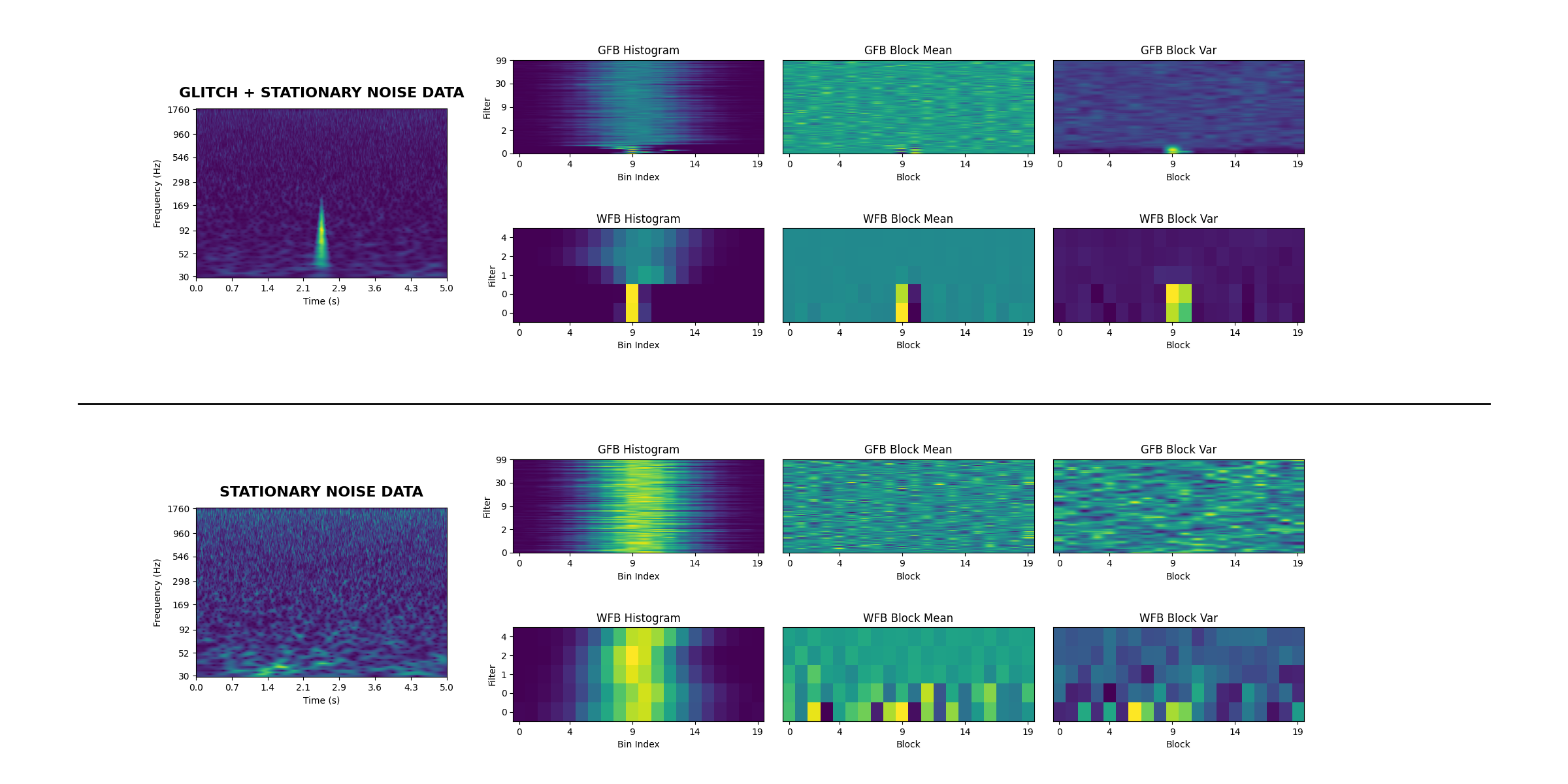}
    \caption{Examples of the features generated from the Gabor filter bank (GFB) and the wavelet filter bank (WFB). The figure is divided into two sections separated by the horizontal line in the middle: the top section pertains to data containing a short-duration glitch with SNR=$72$ added to stationary Gaussian noise, and the bottom is for data with only stationary Gaussian noise. In each section, the leftmost plot is the Q-transform of the data. The panels to its right are organized in a 2$\times$3 grid displaying the three statistical features used in \algoname: histograms (leftmost column), block-wise means (middle column), and block-wise variances (rightmost column). The top and bottom rows of each grid show features obtained from the GFB and WFB, respectively. The axis labeled ``filter" in each panel shows the index of the filter in the corresponding filter bank. 
    The glitch produces patterns in feature space that are distinct from those for noise alone. Note that the GFB matrices shown here are reshaped before being fed to \algoname.}
    \label{fig:features_example}
\end{figure*}

Fig.~\ref{fig:features_plan} presents a schematic diagram of our feature generation pipeline, illustrating how raw time-series data is processed through the GFB and WFB, followed by feature generation and, in the case of the GFB, reshaping.
Fig.~\ref{fig:features_example} illustrates the input features (histogram, block-wise mean, and block-wise variance) obtained for stationary noise versus data containing an anomaly across the GFB and WFB.  This visualization shows how our statistical approach transforms random noise realizations into a structured representation where anomalies can be detected as deviations from expected patterns.
The feature generation described above fundamentally transforms the learning task: instead of attempting to directly reconstruct a noise time series, our autoencoder learns to reconstruct more stable statistical fingerprints of the noise.

\section{Mock training and test datasets}
\label{sec:datasets}

The development of an AI model requires a train-validate-test cycle, where data is partitioned into three distinct sets serving different purposes. The training set is used for the optimization of the model parameters, the validation set guides hyperparameter tuning and prevents overfitting, and the test set provides unbiased evaluation of the final model performance. 

A key aspect of our approach is the exclusive use of simulated Gaussian noise for both training and validation. The performance of the trained model is quantified through receiver operating characteristics (ROC) curves, which characterize the trade-off between detection efficiency and false alarm rate across various detection thresholds. For estimating ROCs, we use a mock dataset comprised of both non-anomalous and anomalous data, the former used for estimating the false alarm and the latter for detection probabilities. 

Under ideal conditions with operating parameters kept fixed, the noise in a GW interferometer should be Gaussian and stationary since that is the nature of the fundamental noise sources such as thermal and photon shot noise. However, real data shows variation in the noise characteristics over $\sim \mathcal{O}(10)$~s to minute timescales\cite{kumar2025parameter}. Hence, simulating realistic data in a GW detector should allow for slow variations in the locally estimated PSD. As such, the simulated noise in the data fed to \algoname uses a design PSD incorporating random perturbations. To better distinguish the simulated data realizations obtained from this model from realizations of stationary noise with a fixed PSD, we use the term mock non-anomalous data in the rest of the paper. Conversely, noise realizations drawn from this model with added simulated anomalies are called mock anomalous data.

\subsection{Training and validation dataset}
\label{sec:train_data}
\begin{figure}
    \includegraphics[width=\linewidth]{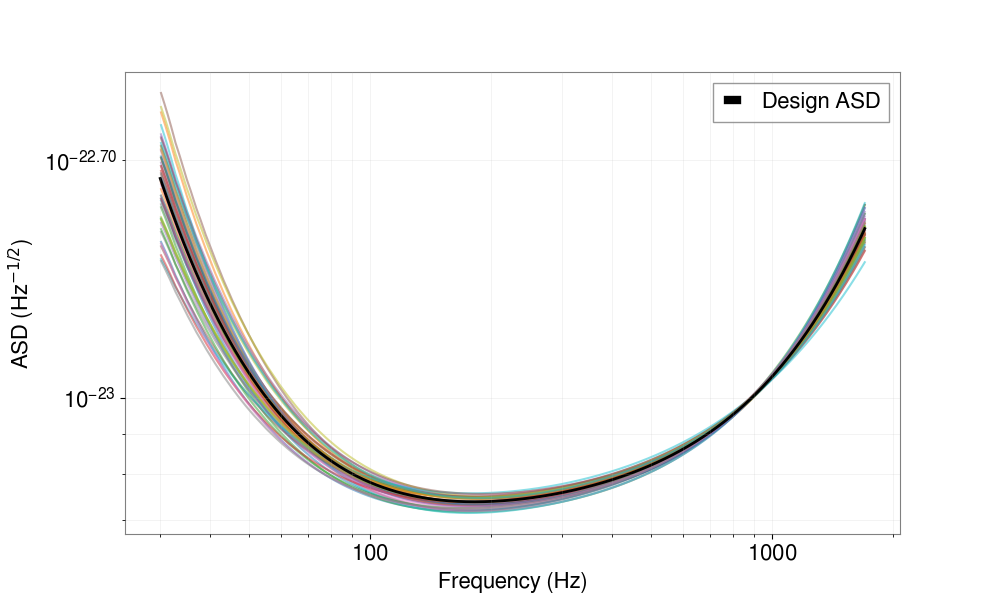}
    \caption{Illustration of perturbed Amplitude Spectral Densities (ASD = $\sqrt{\rm PSD}$), shown by the colored curves, around the base Advanced LIGO Zero-Detuned High Power design sensitivity curve shown in black.}
    \label{fig:perturbed_asd}
\end{figure}

The PSD variation is implemented by first fitting the Advanced LIGO Zero-Detuned High Power design sensitivity one-sided PSD from \texttt{PyCBC}\cite{pycbc2022} as a function of log-frequency with a fifth-order polynomial, followed by perturbing the fitted parameters within $\pm2\%$ of their best-fit values.  Fig.~\ref{fig:perturbed_asd} illustrates this process, showing that the perturbed Amplitude Spectal Densities (ASDs) maintain the essential features of the original sensitivity curve while incorporating variations. 

For the training dataset, we generate $2000$ distinct perturbed PSDs, with one noise realization synthesized per PSD. For the validation dataset, we generate $1000$ additional distinct perturbed PSDs with their corresponding noise realizations. The generation of noise for each perturbed PSD uses frequency domain coloring of $4096$~sec of white Gaussian noise. To avoid edge effects that could contaminate the data, we extract only the central $3600$~sec of each resulting time series.  The validation set serves two purposes: during training, it is used to monitor model convergence and implement an early stopping criterion, and during testing, it is used for false alarm rate estimation.  


\subsection{Test datasets}
\label{sec:test_data}

The test dataset is comprised of the validation dataset and different versions of a mock anomalous dataset, with the versions differing in the type of SNR distribution of the anomalies. The mock anomalous dataset is constructed by adding a simulated signal $\overline{s}$ to $\overline{n}$, a realization of stationary noise. We created a total of $1600$ such realizations in each version, comprising $200$ distinct instances for each of the eight signal classes described in Appendix~\ref{sec:synth_anomalies}. The first five classes, labeled A to E, resemble the classes of short duration glitches identified in Gravity Spy. The next two are chirp signals and the final class is that of IMRPhenomD BBH waveforms. For each data realization, we generated the noise realization from a unique perturbed PSD that was not used in the training or validation datasets. 

\begin{figure*}
    \centering
    \includegraphics[width = \textwidth]{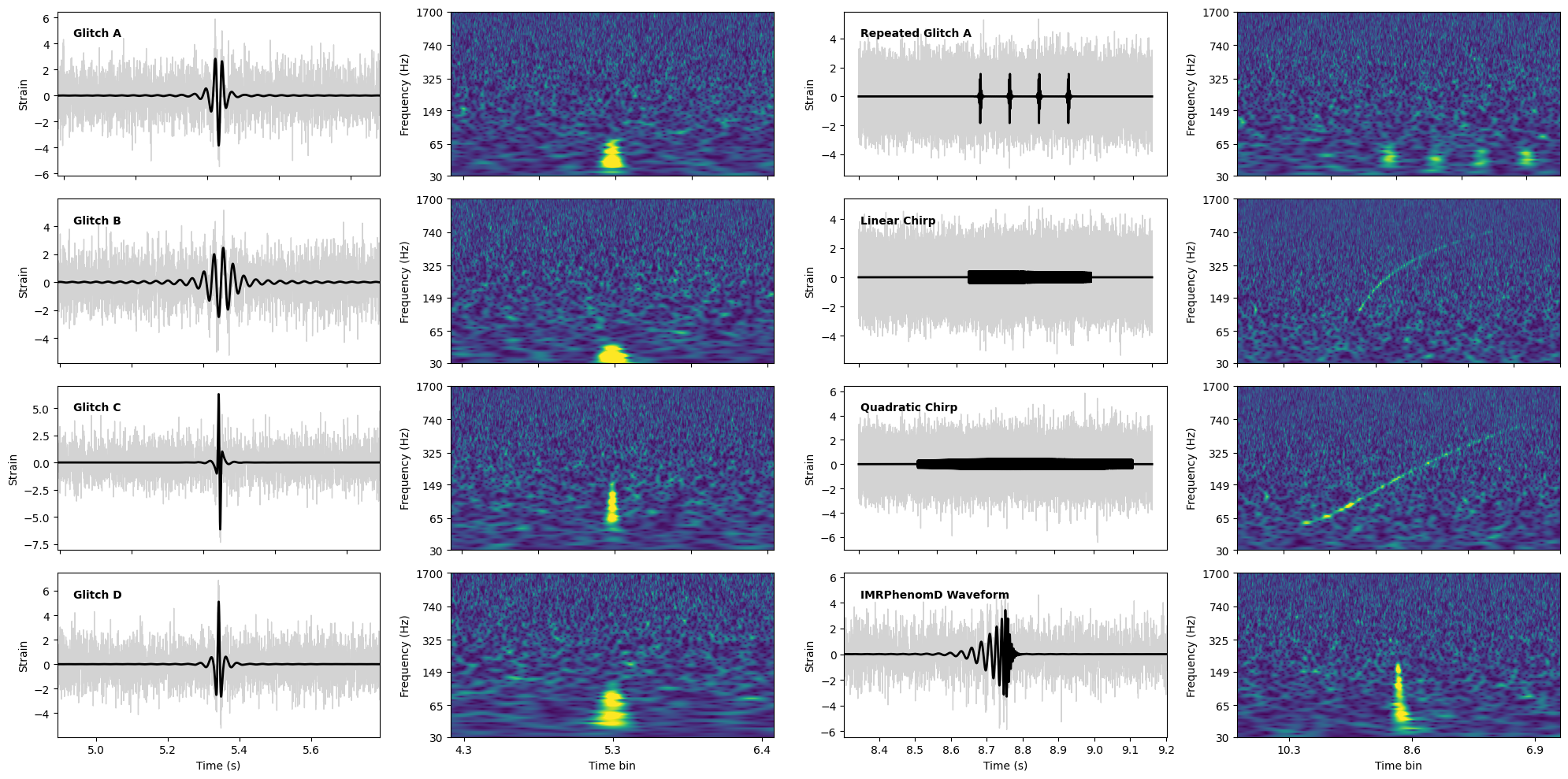}
    \caption{Examples of the eight signal classes, with ${\rm SNR} = 30$, used in the test datasets. Each pair of plots shows the whitened time series (left) and Q transform (right) for one signal type. The first five classes are glitches: Types A, B, C, and D (left column), and Type E (top right), which consists of repeated Glitch A instances. Glitches A and B show oscillatory patterns, Glitch C displays a sharp transient, and Glitch D exhibits a complex structure. The middle right panels show linear and quadratic chirp signals. The bottom right panel presents an IMRPhenomD BBH waveform.}
    \label{fig:anomaly_examples}
\end{figure*}

Each signal is normalized to have a specified signal-to-noise ratio (SNR) with respect to the PSD of the noise it is added to. For a signal $\overline{s}$, the SNR is the one defined for matched filtering, namely,
\begin{equation}
    {\rm SNR} = \sqrt{\sum_{i \in \mathcal{F}} \frac{\tilde{s}^*_i \tilde{s}_i}{S_{n,i}} \Delta f}\;,
\end{equation}
where $\tilde{s}_i$ denotes the discrete Fourier transform (DFT) of the signal at the $i$th DFT frequency, $S_{n,i}$ is the two-sided noise PSD used for the specific data realization,  $\Delta f$ is the frequency resolution of the DFT, and $\mathcal{F}$ is the set of  positive and negative DFT frequencies. Fig. \ref{fig:anomaly_examples} shows representatives from each class of simulated anomalies.

Among the different versions of the test dataset used for evaluating performance, one version uses SNRs drawn from a log-uniform pdf over the range $[10, 20]$, designed to explore performance within this specific intermediate SNR band. The other versions utilize fixed SNR injections with ${\rm SNR}\in (5, 10, 15, 20)$ to characterize detection performance at specific signal strengths.


\section{DNN model and training}
\label{sec:dnn_model}


The feature matrices described in Sec.~\ref{subsec:features} carry information that is sequential across both time and frequency and correlated along each dimension. Among DNN models suitable for such sequential data, 
the transformer architecture~\cite{vaswani2017attention} has rapidly gained widespread use. This success is due to the paradigm introduced in transformers based on `attention mechanisms' that has proven highly effective for capturing long-range dependencies in sequential data such as natural language.
The key innovation is `self-attention'. Conceptually, when the model processes one part, called a token, of the sequence, self-attention allows it to look at other tokens and score them on their relevance to the current token. Based on these scores, it computes a weighted representation, emphasizing information from the most relevant tokens, allowing it to learn complex dependencies regardless of their distance in the sequence.


For applying the transformer model in our case, the feature matrices must be tokenized in the best way. 
In the standard application of Transformers to multivariate time series, the focus is typically on the time dimension with the feature vector at each time step, i.e., a column of the matrix, treated as a single token.
We follow a different approach in which the encoder adapts the iTransformer architecture that was specifically proposed in~\cite{liu2023itransformer} for multivariate time series where inter-variable relationships are of critical importance. This is true in our case since a broadband glitch or chirp can appear across multiple rows of the feature matrix whereas the majority of the columns would just be related to noise. 

For our input feature matrices, the iTransformer approach involves treating each row as an individual token. 
This feature-centric tokenization enables the model to learn correlations and dependencies between the temporal patterns of features across different frequency bands through self-attention. Although the GFB feature matrix is reshaped, the rows still contain features from different frequency bands and the same  iTransformer mechanism applies to the identification of dependencies across frequency.  


\subsection{Model architecture}
\label{sec:model_arch}
\begin{figure}
    \centering
    \includegraphics[width=\columnwidth]{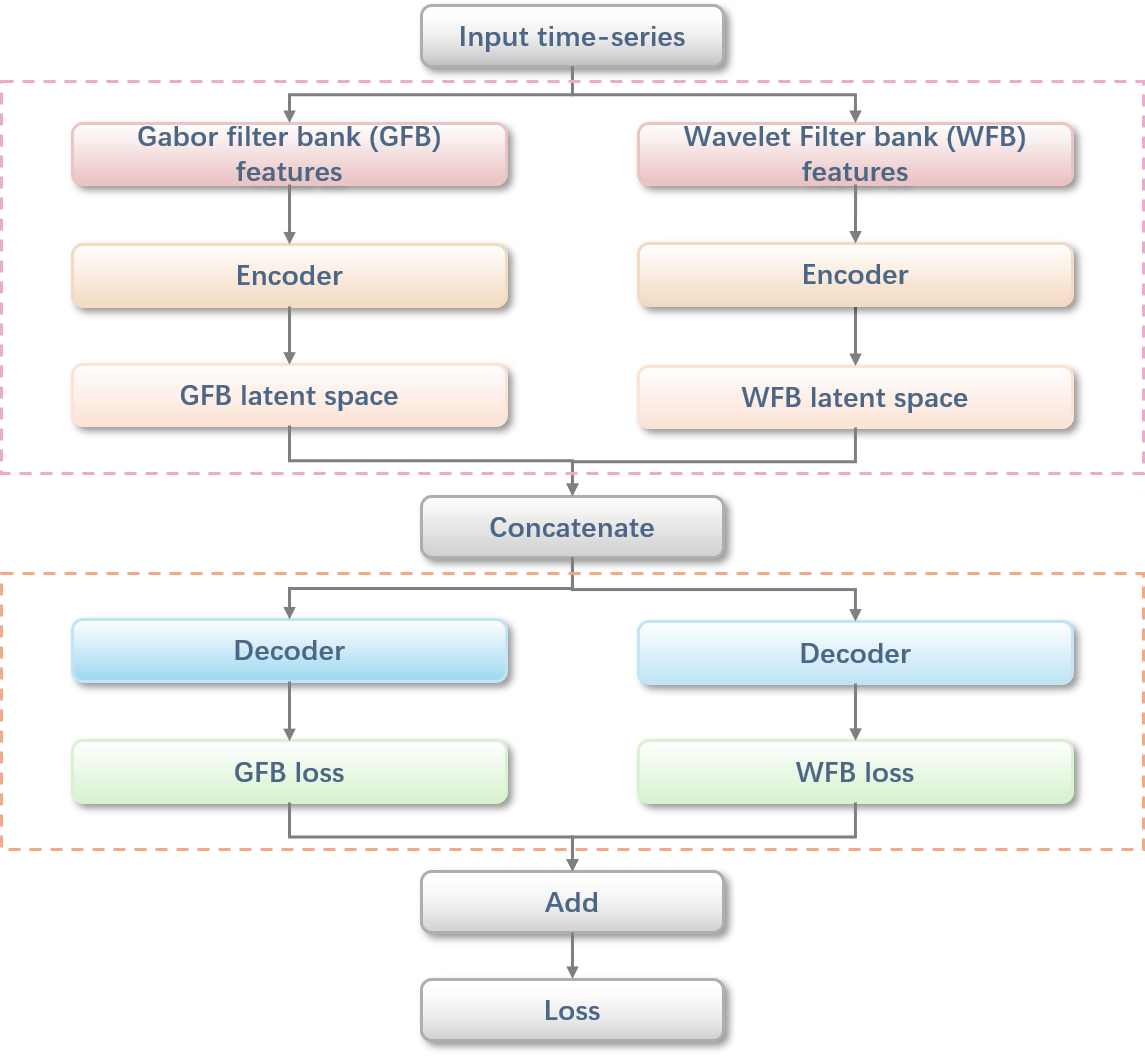}
    \caption{Architecture of the ABNORMAL model. The diagram illustrates the dual-path autoencoder design that processes GFB and WFB features through parallel encoder paths. Both paths extract complementary time-frequency information  before combining in a shared latent space for multi-view fusion. The decoder then reconstructs the features, with dedicated output heads for each feature type.}
    \label{fig:abnormal-model}
\end{figure}

Guided by the above considerations, \algoname employs a dual-path autoencoder architecture designed for multiview feature processing as shown in Fig.~\ref{fig:abnormal-model}, with iTransformer based components in both the encoders and decoders. The network processes the GFB and WFB features through parallel paths that combine in a shared latent space before reconstruction. This design enables capturing complementary time-frequency information at different resolutions while maintaining separate processing streams optimized for each feature type. 

Each encoder has three iTransformer layers. A single iTransformer layer is composed of an $8$-head self-attention mechanism and a feed-forward network (dimension $2048$). The overall model architecture uses a dimension of $512$, a dropout rate of $0.1$, and ReLU activation.
The latent features produced by the two encoder paths are concatenated and sent to the two decoder paths. (It is the concatenation step that necessitates reshaping the GFB matrix to align feature dimensions with those of the WFB path.) Each decoder maps the combined latent representation, corresponding to both GFB and WFB, back to the dimensions of the original feature matrix for its respective output.

\begin{figure*}
    \centering
    \includegraphics[width=\textwidth]{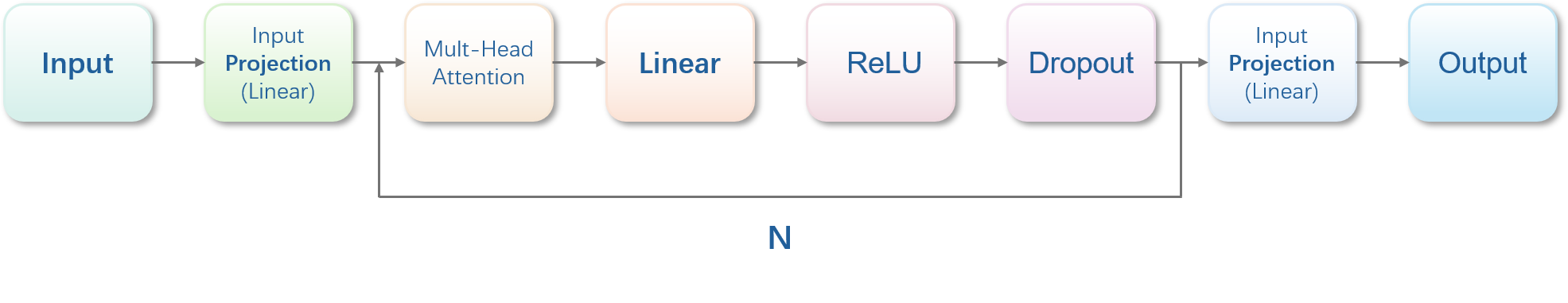}
    \caption{Architecture of the modified decoder in \algoname. The block consisting of Multi-Head Attention , a Linear transformation, ReLU activation, and Dropout is repeated N = 2 times. This is followed by an input projection and an output projection. }
    \label{fig:decoder-architecture}
\end{figure*}

The decoder, shown schematically in Fig.~\ref{fig:decoder-architecture}, enhances the standard iTransformer architecture by replacing its simple output projection with a multi-layer network. This network processes the combined latent representations from the shared space before reconstruction and follows a sequential architecture: an input projection (linear), a multi-head attention block (with 8 heads), a linear transformation, a ReLU activation function, a dropout layer (rate 0.1), and a final output projection (linear). Separate output heads then reconstruct the GFB and WFB features, preserving path-specific temporal and frequency characteristics while enabling information sharing through the latent space.

\subsection{Model training}
\label{sec:model_train}
The training process of a DNN is accomplished in several consecutive passes (called epochs). The  performance of the model is evaluated using a loss function after each pass and the gradients of the loss with respect to the model parameters (the weights and biases of the artificial neurons) are computed. These gradients are then used to update the model parameters in a direction that reduces the loss, typically using an optimization algorithm.

The model is optimized with respect to a single objective: minimizing the total reconstruction error, denoted as the total loss $\mathcal{L}_{\text{total}}$. For clarity in explaining the model's objective and its staged training, we define this total loss as the sum of two components, $\mathcal{L}_{\text{GFB}}$ and $\mathcal{L}_{\text{WFB}}$, which measure the reconstruction errors for the GFB and WFB paths, respectively:
\begin{eqnarray}
        \mathcal{L}_{\text{total}} &=& \mathcal{L}_{\text{GFB}} + \mathcal{L}_{\text{WFB}} \nonumber \\
        &=& \sum_{p \in \{\text{GFB}, \text{WFB}\}} \!\!\! \!\!\!\! \!\!\!\! (N_f N_t^p)^{-1}  \sum_{i=1}^{N_f} \sum_{j=1}^{N_t^p} (X^p_{ij} - \widehat{X}^p_{ij})^2\;,
\end{eqnarray}
where $X^p$ is the feature matrix from path $p$, $\widehat{X}^p$ is its reconstruction, and $N_f$ and $N_t^p$ are the matrix dimensions. During the detection phase, this same $\mathcal{L}_{\text{total}}$ serves as the anomaly indicator, with higher values indicating larger deviations from stationary noise behavior.

While the overall loss function is unified, the training is conducted in stages to ensure stability. In the initial phases, we alternately "freeze" the parameters of one path to train the other. This staged approach allows each path to learn a stable representation of its specific features before they are fine-tuned jointly. The training phases and their corresponding epochs are as follows
\begin{itemize}
  \item Epochs 1-20: The parameters of the GFB path are frozen, and only the WFB path parameters are updated (learning rate = $5\times10^{-4}$).
  \item Epochs 21-70: The parameters of the WFB path are frozen, and only the GFB path parameters are updated (learning rate = $1\times10^{-4}$).
  \item This alternating pattern is repeated until epoch 140.
  \item Epochs 141-300: All parameters in the network are unfrozen, and both paths are trained simultaneously (learning rate = $1\times10^{-4}$).
\end{itemize}
In each training phase, the model parameters are optimized using the Adam algorithm. To prevent unstable learning behavior, gradient clipping with a maximum norm of $1.0$ is applied, which constrains the gradients and mitigates the risk of excessively large parameter updates. For each epoch, we also compute the loss for the validation dataset described in Sec.~\ref{sec:train_data} and track its progression over time. To reduce the impact of statistical fluctuations, validation loss is averaged over a rolling 10-epoch window. To prevent overfitting to the training data, the training is terminated when the smoothed validation loss shows no significant improvement over $50$ consecutive epochs after the initial training phases. The final model parameters are selected based on achieving the minimum reconstruction error on the validation set over the epochs following the start of the simultaneous training phase of both paths (i.e., from epoch $141$ onwards). For the model used in this paper, termination happened at epoch number $205$ and epoch $197$ yielded the lowest validation loss.

\section{Considerations for real data}
\label{sec:results_threshold}
\begin{figure*}
    \centering
    \includegraphics[width=\textwidth]{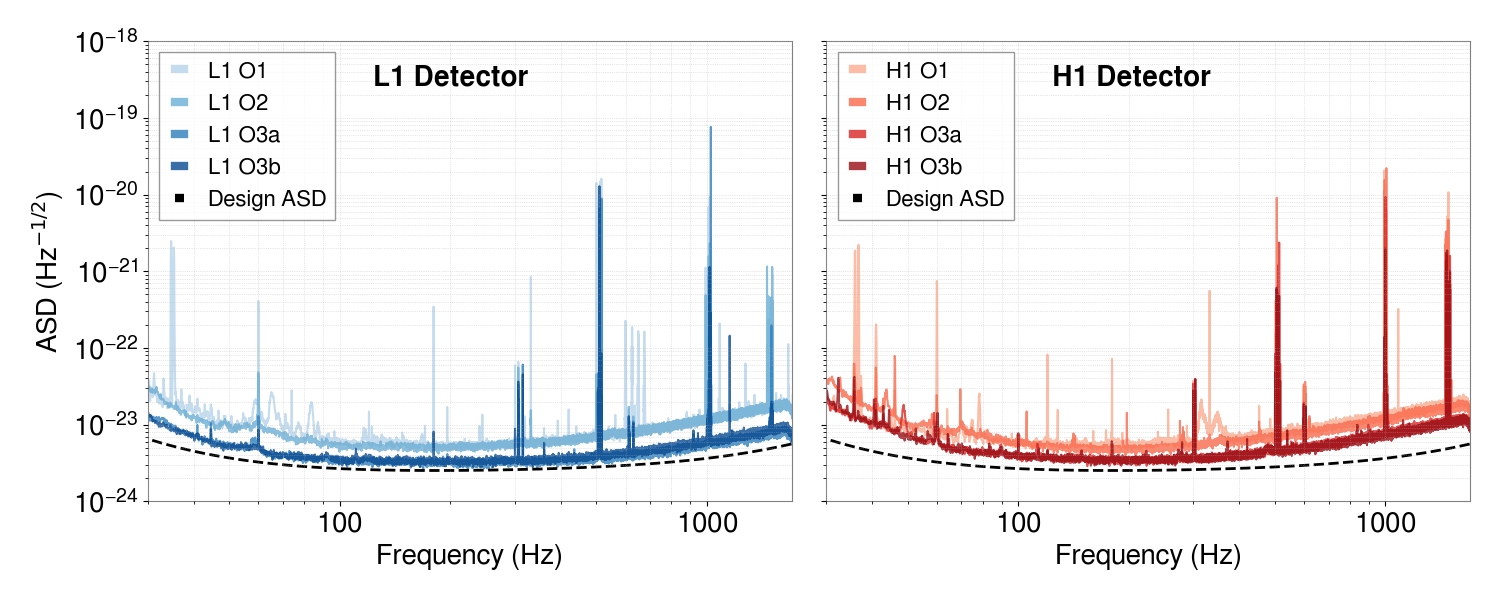}
    \caption{Representative ASDs for L1 and H1 across O1, O2, O3a, and O3b. These are compared against the design sensitivity curve (dashed black line), which illustrates both the variation in noise characteristics between different observing runs and detectors, and the significant deviations caused by lines (sharp spectral features).}
    \label{fig:orun_psds}
\end{figure*}

 \algoname is trained exclusively on mock non-anomalous data as described in Sec.~\ref{sec:train_data}. However, as shown by the representative ASDs of LIGO-Hanford (H1) and LIGO-Livingston (L1) in Fig.~\ref{fig:orun_psds}, there are significant features in the real data that are not captured in the mock data generation. In particular, real data includes high power narrowband noise features, commonly called lines, that dominate its time domain behavior. As a result, the loss distribution for real non-anomalous data deviates significantly from that obtained from mock data. This affects the false alarm probability corresponding to a given detection threshold for anomalies, making the loss distribution obtained from mock data unusable for this purpose.
 
Bringing the two distributions into better alignment requires the removal of strong lines from real data. Since the line removal method we use also alters the PSD of the residual, the mock non-anomalous loss distribution must be recomputed before estimating an anomaly detection threshold. An alternative approach would be to train \algoname on mock data that includes line noise. However, this is technically more challenging because lines need not have predictable amplitude and frequency modulations, and simulating data with a large dynamic range is subject to numerical issues, with no assurance of achieving the desired result.

In the following subsections, we describe the above process in more detail, starting with line removal, choosing a well-behaved segment of (line removed) real data for PSD estimation, and generation of mock non-anomalous data for obtaining the loss distribution and recalibrating the detection threshold. It is important to emphasize here that \algoname itself is not retrained at any point in the above process, only the loss distribution is recomputed. 

\subsection{Line removal}
\label{subsec:preprocessing} 

Our line removal process consists of two main steps: line identification and notch filtering.  Since we need to address only loud lines, the former step can be carried out with any sufficiently long stretch of (science mode) real data that is visually devoid of exceptionally loud transient anomalies. We use data segments of equal lengths with the highest quality (as given by the DQ flags) sampled at 4096~Hz from each of the observing runs up to O3b. (This resulted in a segment length of $8191$~sec.) In all cases, the amplitude spectral density (ASD) for line identification is estimated with a $0.1$~Hz frequency resolution and only lines in $[30,1700]$~Hz are considered.  

\begin{figure*}
    \centering
    \includegraphics[width=\textwidth]{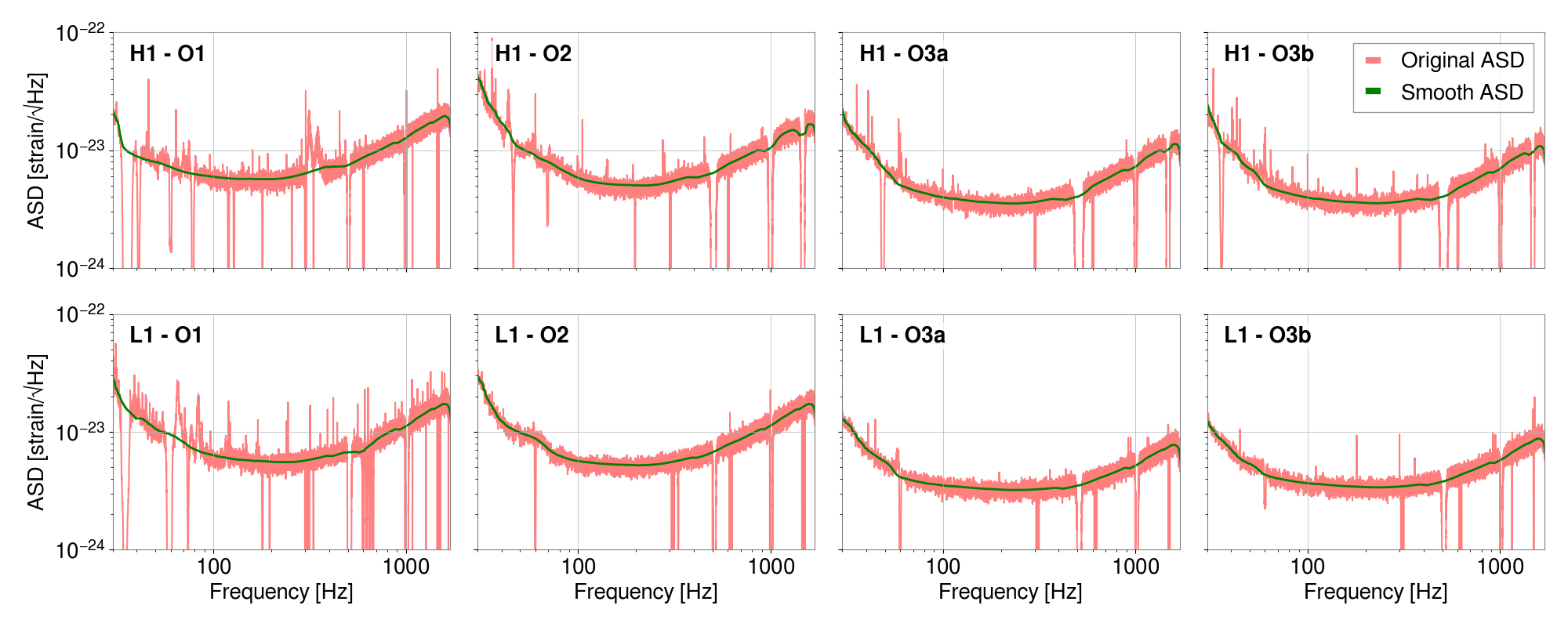}
    \caption{Amplitude Spectral Density (ASD) across different detectors and observing runs after line removal. The top and bottom rows are ASDs corresponding to H1 and L1, respectively, from data segments belonging to (left to right) O1, O2, O3a, and O3b. Each panel shows the ASD (red) after line removal and the running median estimate (green) used for mock matching data generation (c.f., Sec.~\ref{sec:matchmockdata}). The vertical axis range has been restricted for visual clarity.}
    \label{fig:smooth_ASD}
\end{figure*}

A semi-automated process is used for identifying lines and training the notch filters. The details of the line identification method and the notch filters are provided in Appendix~\ref{sec:ln_rmv_method}. The result of line removal is shown in 
Fig.~\ref{fig:smooth_ASD} along with the running median smoothed ASD that is used in the next step.  

\begin{figure*}
    \centering
    \includegraphics[width=\textwidth]{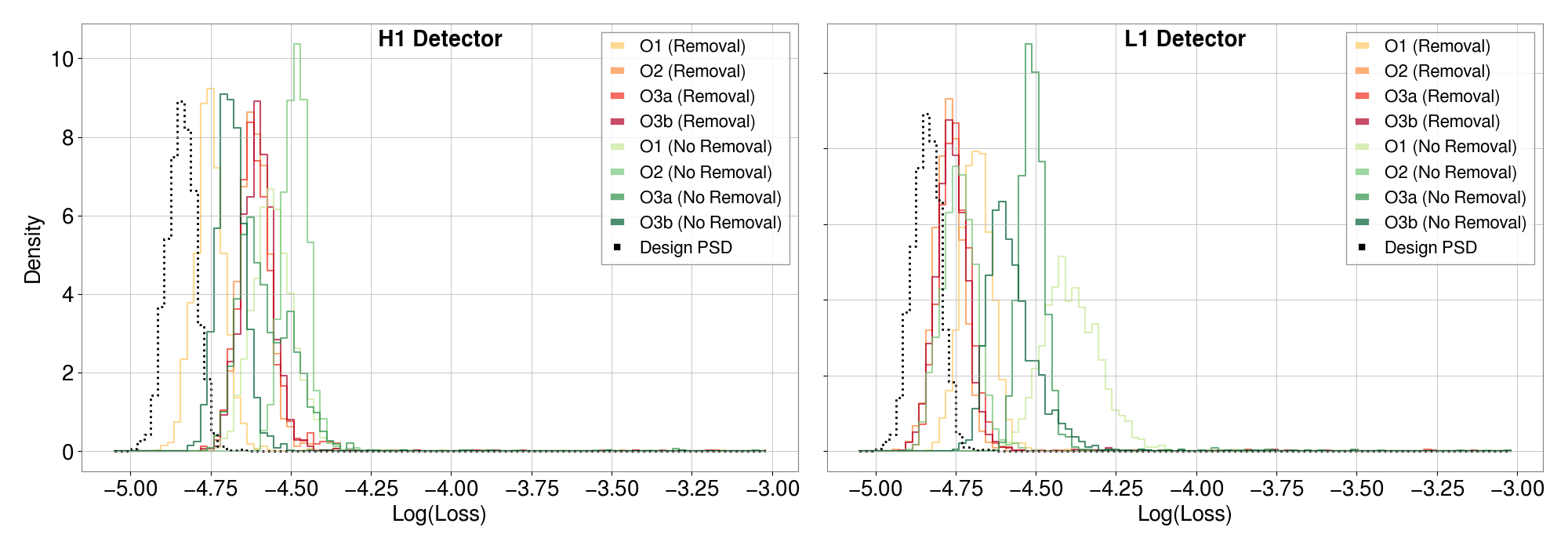}
    \caption{Impact of line removal on the loss distributions for different observing runs. Each panel shows the distribution of  loss values for data with and without line removal and compares them with the loss distribution (dotted curve) obtained from mock non-anomalous data. The distributions have a significant shift toward lower loss values after line removal.}
    \label{fig:with_without_lineRemoval_loss}
\end{figure*}

Fig.~\ref{fig:with_without_lineRemoval_loss} shows the significant impact of line removal on the loss distribution obtained from real data. It is seen that the removal of lines brings the loss distribution of real data much closer to that based on the design PSD. However, the alignment is not complete, and part of the reason is that the PSD of real data after line removal has deep notches in it that again deviate from the design PSD. To get better alignment, the mock non-anomalous data must have the same type of notches in the spectral domain. Thus, it must be regenerated to better match the spectral notches in real data PSD. Data generated in this manner is called matching mock data in the following to emphasize that it is  derived from real data. It is also understood from here on, unless stated otherwise, that all real data segments from a given run are subjected to line removal using the notch filters trained for that run.

\subsection{Matching mock data generation}
\label{sec:matchmockdata}

The PSD used for generating matching mock data, and its associated loss distribution, should be estimated from a real data segment that is minimally contaminated by long-lasting anomalies. To identify such data, we use \algoname itself on real data using the following procedure. 
For each observation run, we select the longest continuous data segment without gaps and
divide it into overlapping subsegments of $1000$~sec duration, with $50\%$ overlap between adjacent subsegments. Each subsegment is processed through \algoname, using $N_{\text{blks}}=20$, to obtain a loss value. 
The subsegment with the lowest loss value, is then used for estimating the PSD needed for generating matching mock data.  
The PSD for this purpose is estimated with a frequency resolution of $0.05$~Hz, which differs from the coarser resolution used in the line identification process.
Note that, here, the time-scale of anomalies that \algoname is sensitive to is set by the above choice of $N_{\rm blks}$ to $\gtrsim 200$~sec. This is because we need only guard against very long-duration anomalies that can significantly bias the estimated PSD floor.

The estimated ASD, which contains spectral notches due to line removal, undergoes a running median smoothing that uses frequency-dependent block lengths: primarily $25$~Hz (corresponding to 1000 frequency bins at a resolution of $1/40$~Hz) and $112.5$~Hz (4500 bins), transitioning between them across the spectrum. The running median estimates shown in Fig.~\ref{fig:smooth_ASD} show that this procedure is effective at smoothing over the notches. Using the smoothed ASD as the target power spectral density, we proceed to generate synthetic stationary Gaussian noise. The generated noise realization then undergoes the same line removal process as the real data, using identical notch filter parameters. After removing edge effects from filtering, we obtain a realization of matching mock data. 

\subsection{Detection threshold estimation}
\label{sec:detection_thresholds}
Having set up the procedure for generating matching mock data, one can generate as large a stretch of it as needed for estimating detection thresholds and false alarm rates. 
The generated data is segmented, with the number of segments decided by the false alarm rate at which the detection threshold needs to be estimated,  and the \algoname loss values obtained for each segment. 
In this paper, matching mock data generated for each observing run is divided into 5-second segments, with $1003$ samples overlapped between consecutive segments for edge effect removal, aligning with the real data analysis pipeline. In total, $5000$ such 5-second segments are generated. 

\begin{figure*}[htbp]
    \centering
    \includegraphics[width=1\textwidth]{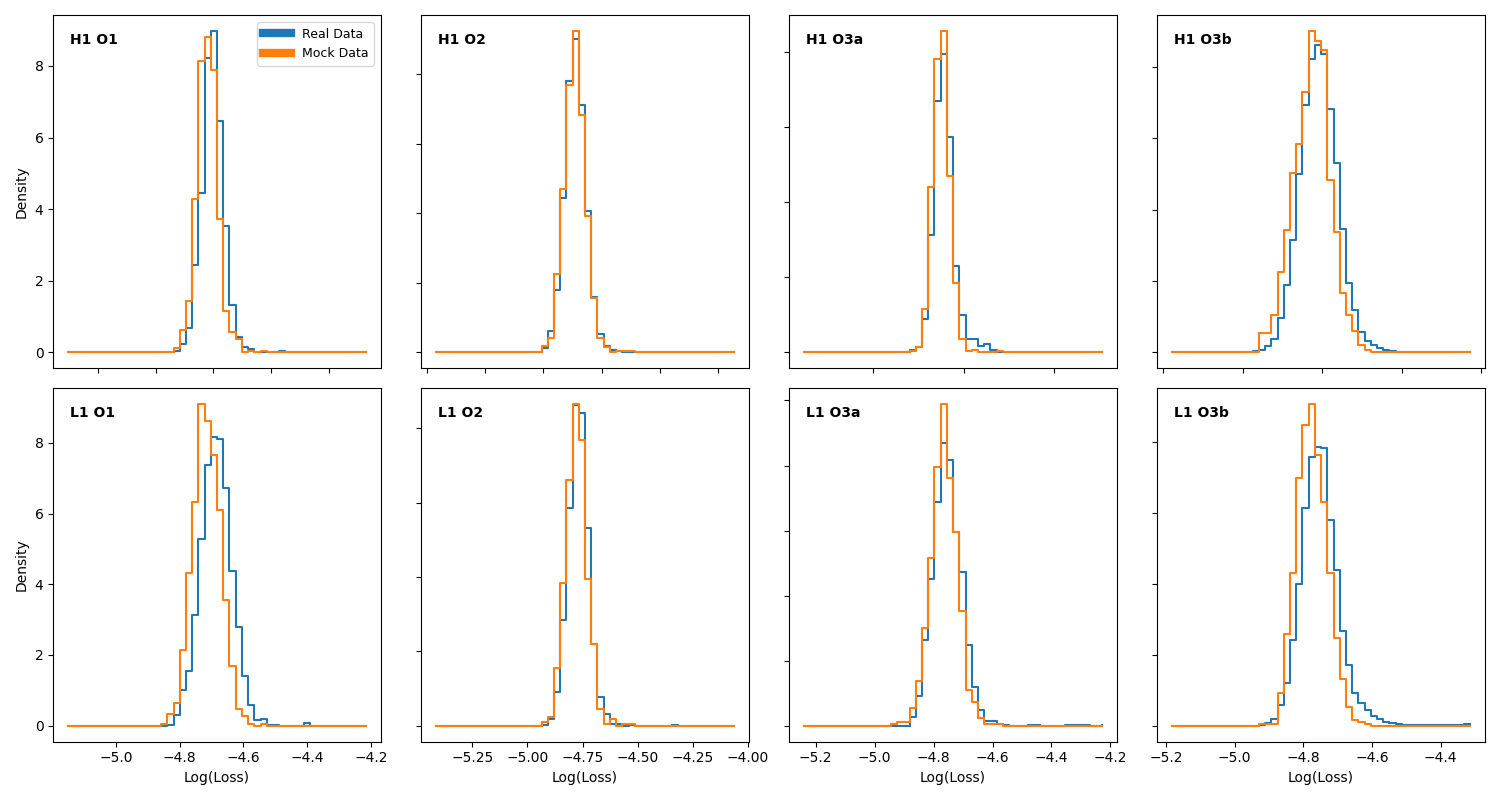}
    \caption{Comparison of loss distributions between real GW data (blue) and corresponding mock data (orange) for LIGO Hanford (H1, top row) and LIGO Livingston (L1, bottom row) detectors during observing runs O1, O2, O3a, and O3b. The results were obtained using $\nblks=20$ and a segment length of $5$~sec. The close alignment between real and mock data distributions, particularly in the low-loss regime (corresponding to more stationary data), validates our approach of using preprocessed mock data for anomaly detection threshold determination.}
    \label{fig:loss-real-mock}
\end{figure*}

Figure~\ref{fig:loss-real-mock} compares loss distributions from  real data segments  and their corresponding matching mock data (analyzed in $5$~sec intervals) for all observation runs and detectors. For any given observation run from a well-behaved detector, one expects the mode of the loss distribution to correspond to the majority of the data being free of significant anomalies. Hence, if the matching mock data is representative of stationary real data, the modes of the loss distribution are expected to match. The fact that this alignment is achieved for most of the observation runs lends support to the procedure for generating matching mock  data.
There exist slight misalignments between the modes of the loss distributions for the L1-O1 and L1-O3b runs. For L1-O1, this discrepancy is
likely attributable to the higher number of spectral lines (c.f., Fig.~\ref{fig:smooth_ASD}), that were not fully addressed by our semi-automated line removal process. 
We do not have a clear explanation yet for the observed discrepancy for L1-O3b. 
 
The obtained distribution of loss values is compared with that from the real data associated with the matching mock data and the former is shifted to align its mode with that of the latter. The detection threshold for a desired false alarm probability is then read off from the shifted matching mock data loss distribution. In this paper, we use thresholds corresponding to a False Alarm Probability of $\approx 0.5\%$, which is equivalent (given the $5$~second segment length) to a False Alarm Rate (FAR) of $\approx 10^{-3}$~events/sec, or an inverse FAR of $\approx 1$~event in $16$~minutes.

\section{Results}
\label{sec:results}
Our results are presented in two parts. First, we evaluate the performance of \algoname using the mock datasets described in Sec.~\ref{sec:datasets}.
Next, using the detection thresholds obtained as described in Sec.~\ref{sec:detection_thresholds},
we analyze LIGO data from multiple observing runs and present (i) the performance of \algoname on a subset of observed GW signals, and (ii) examples from the diverse landscape of short and long time scale anomalies. The number of blocks is the same for both the short and long time scale analyses, $\nblks = 20$, but the segment lengths are set to $5$~sec and $20$~sec, respectively. All analyses of real data use the detection thresholds obtained in Sec.~\ref{sec:results_threshold} with a FAR of $\approx 10^{-3}$~events/sec.

\subsection{Performance on mock data}
\label{sec:results_mock}

\begin{figure*}[t]
    \centering
    \includegraphics[scale=0.4]{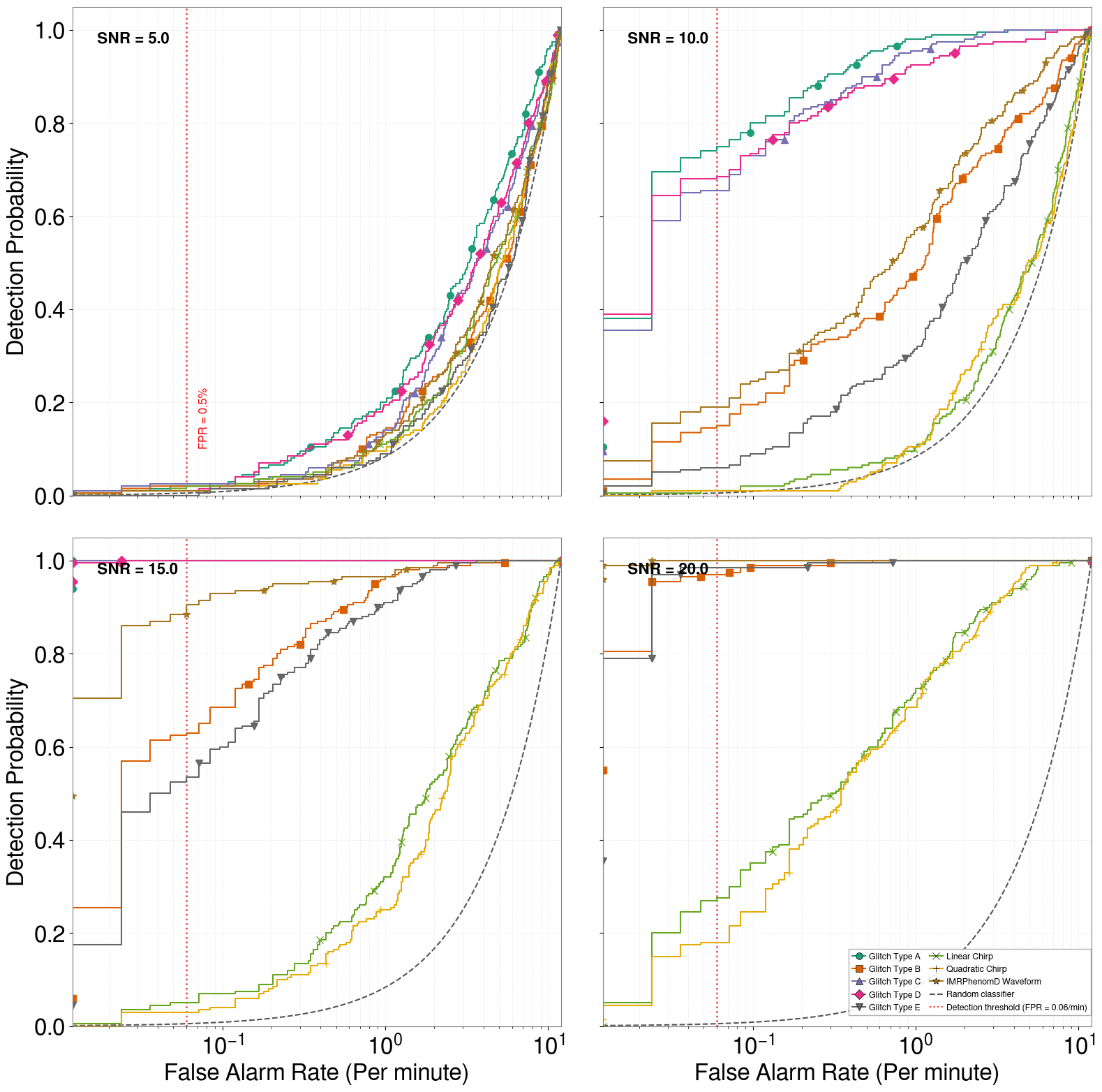}
    \caption{ROC curves for the synthetic anomaly classes described in Appendix~\ref{sec:synth_anomalies} at fixed  ${\rm SNR} \in \{5, 10, 15,20\}$. 
    Each panel corresponds to one value of SNR as noted in its top left corner.
     The correspondence between the different glitch classes and the ROC curves legend is shown in the bottom-right panel.}
    \label{fig:ROC_curves}
\end{figure*}

\begin{figure}[t]
    \centering
    \includegraphics[scale=0.4]{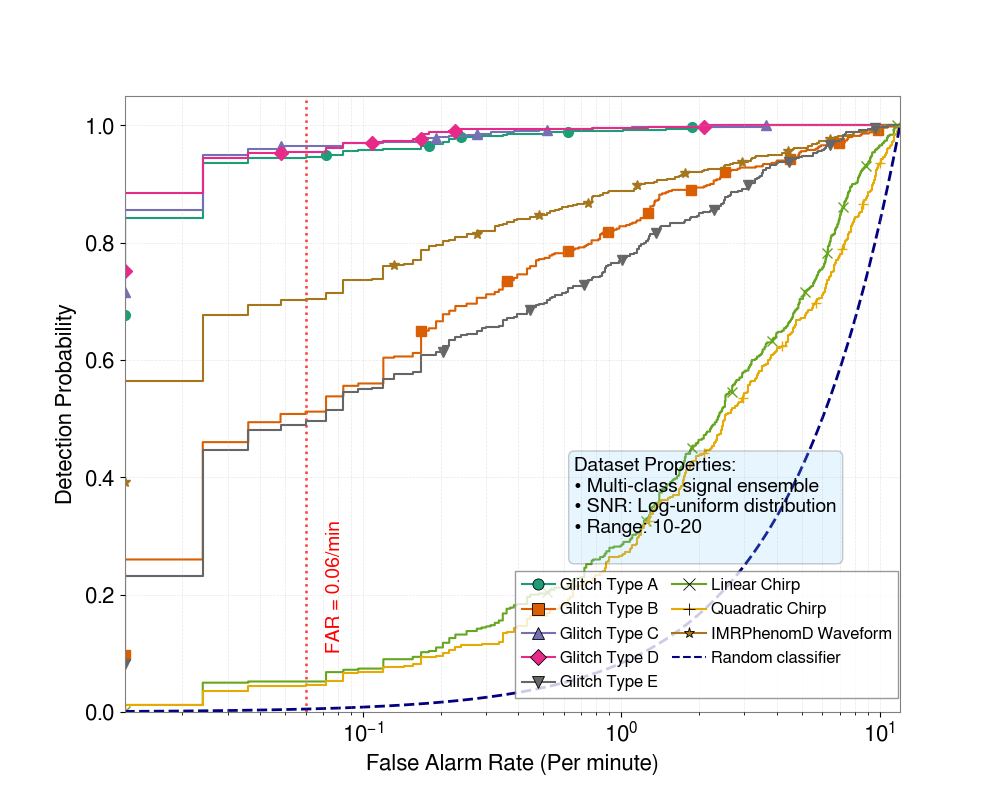}
    \caption{Receiver Operating Characteristic (ROC) curves for different signal classes using an ensemble of signal-to-noise ratios (SNR). The SNRs were sampled from a log-uniform probability density function (pdf) spanning the range $[10, 20]$. The legend identifies the curve for each signal class. }
    \label{fig:ROC_ensemble} 
\end{figure}

We evaluate the performance of \algoname using two standard metrics: the Receiver Operating Characteristic (ROC) curve, which plots the detection probability against the false alarm probability, and the Area Under this Curve (AUC), which provides a single scalar measure of detection performance.  These metrics are computed for the two versions of test datasets described in Sec.~\ref{sec:test_data}. The ROC curves for the  the fixed and variable SNR datasets are shown in Fig.~\ref{fig:ROC_curves}) and Fig.~\ref{fig:ROC_ensemble}, respectively. The latter  complements the former by exploring the performance across SNR values not contained in the discrete set. The AUC values for the fixed SNR cases are given in Table~\ref{table:AUC_table}.
These results show that \algoname has strong detection capability for the short-duration glitches, while showing relatively lower sensitivity to longer-duration signals such as chirps. This performance difference shows that the input features considered in this paper work well for short duration broadband glitches, which constitute the bulk of glitches in Gravity Spy, but need to be enhanced for chirps with more sophisticated time-frequency transforms. 
 
\begin{table}
    \begin{ruledtabular}
    \begin{tabular}{lcccc}
    Waveform & SNR = 5.0 & 10.0 & 15.0 & 20.0 \\
    \hline
    Glitch Type A & 0.671 & 0.990 & 1.000 & 1.000 \\
    Glitch Type B & 0.552 & 0.812 & 0.983 & 0.999 \\
    Glitch Type C & 0.618 & 0.984 & 1.000 & 1.000 \\
    Glitch Type D & 0.634 & 0.973 & 1.000 & 1.000 \\
    Glitch Type E & 0.531 & 0.737 & 0.977 & 0.999 \\
    Linear Chirp & 0.554 & 0.555 & 0.747 & 0.920 \\
    Quadratic Chirp & 0.540 & 0.551 & 0.728 & 0.919 \\
    IMRPhenomD & 0.561 & 0.855 & 0.993 & 1.000 \\
    \end{tabular}
    \end{ruledtabular}
    \label{tab:auc_comparison}
    \caption{AUC values obtained from the ROC curves in Fig.~\ref{fig:ROC_curves} across different SNR and signal classes.}
    \label{table:AUC_table}
\end{table}

Following standard practice in efficiency studies of GW data analysis methods, Table~\ref{table:SNRDet_table} provides a complementary view of performance in terms of the SNRs needed to achieve specified detection probabilities of $90\%$, $75\%$, $50\%$, and $25\%$ at a fixed false alarm rate.
The results demonstrate that short-duration glitch signals are detectable at lower SNR values compared to longer-duration signals, consistent with the ROC analysis. At the $50\%$ detection probability threshold, the short duration glitches as well as the IMRPhenomD signals are detectable at SNR values between $8.28$ and $14.63$, while chirp signals require substantially higher SNR values of $21.77$ to $22.32$.
\begin{table}
    \begin{ruledtabular}
    \begin{tabular}{lcccc}
    Waveform & 90\%  & 75\%  & 50\%  & 25\%  \\
    \hline
    Glitch Type A & 13.000 & 10.000 & 8.284 & 6.575 \\
    Glitch Type B & 18.971 & 16.765 & 13.646 & 11.042 \\
    Glitch Type C & 13.551 & 11.377 & 8.770 & 6.786 \\
    Glitch Type D & 13.413 & 11.032 & 8.630 & 6.778 \\
    Glitch Type E & 19.056 & 17.399 & 14.632 & 12.000 \\
    Linear Chirp & 24.921 & 23.740 & 21.772 & 19.444 \\
    Quadratic Chirp & 26.200 & 24.130 & 22.319 & 20.507 \\
    IMRPhenomD & 14.965 & 13.916 & 12.168 & 10.420 
    \end{tabular}
    \end{ruledtabular}
    \caption{\label{tab:snr_comparison}Required SNR values for achieving 90\%, 75\%, and 50\% detection probabilities across different signal types, at a FAR of $\approx 10^{-3}$~events/sec. }
    \label{table:SNRDet_table}
 \end{table}

\begin{figure*}
    \centering
    \includegraphics[scale=0.45]{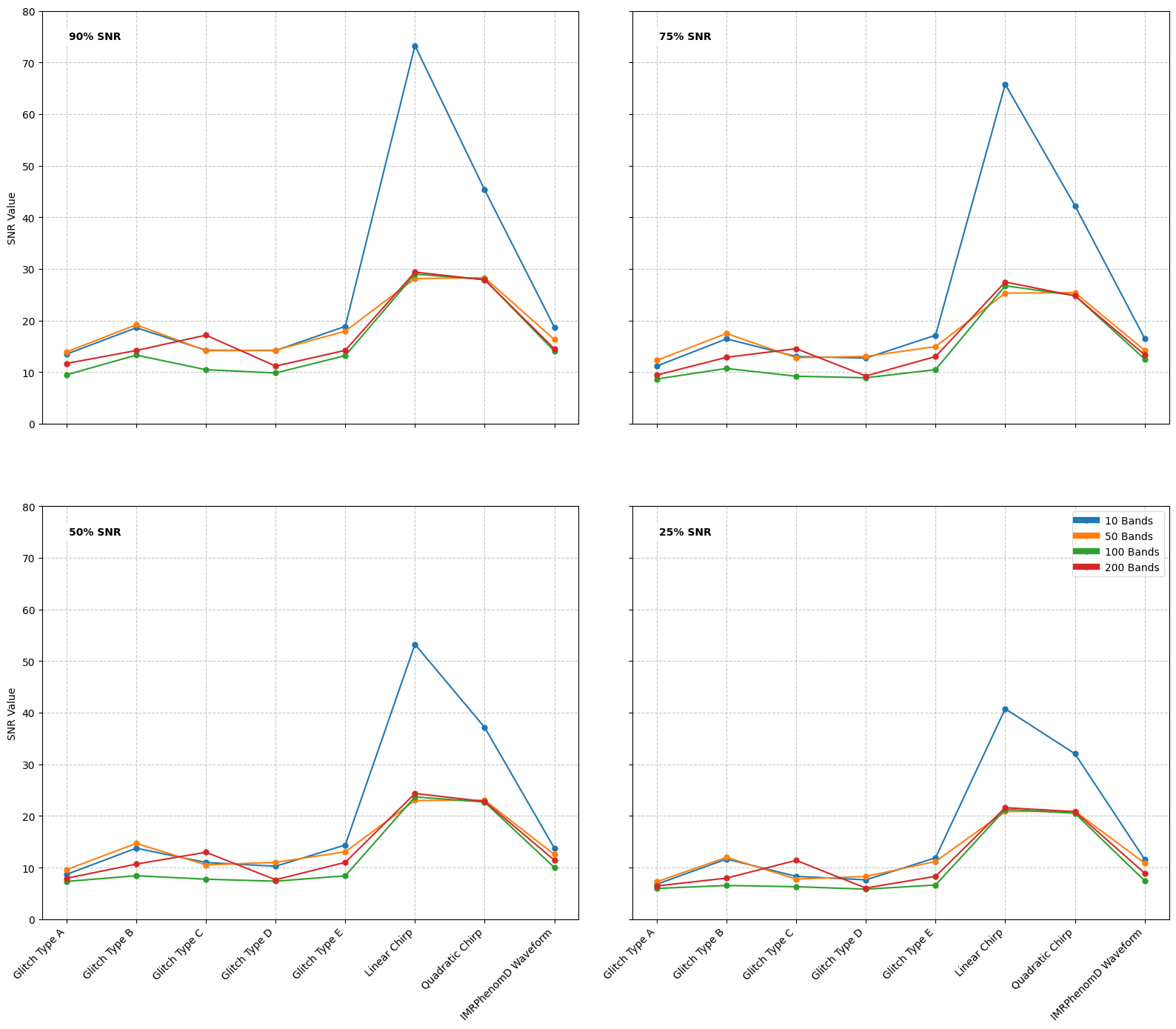}
    \caption{Detection sensitivity of \algoname as a function of the number of filters used for feature generation in the Gabor Filter Bank (GFB). The figure consists of four panels, each corresponding to a specific detection probability, $90\%$, $75\%$, $50\%$, and $25\%$, at a FAR of $\approx 10^{-3}$~events/sec. Within each panel, the y-axis displays the minimum SNR required to achieve that detection rate for the signal types shown on the x-axis. The different curves in each panel correspond to different numbers of GFB filters $(10, 50, 100, 200)$.}
    \label{fig:bandSelection}
\end{figure*}

An important hyperparameter in our feature generation pipeline is the number of filters used for the GFB. For this, we conducted a systematic evaluation using the AUC on our test dataset across different signal types.
Fig.~\ref{fig:bandSelection} shows the SNR values required to achieve different detection rates $(25\%, 50\%, 75\%, 90\%)$ for various signal types when using different numbers $(10, 50, 100, 200)$ of frequency bands. 
Our analysis reveals that increasing the number of filters from $10$ to $100$ consistently improves detection sensitivity across most signal types, particularly for short-duration glitches (Types A, B, C, D and E). However, the performance improvement plateaus and sometimes slightly degrades when increasing from $100$ to $200$ filters, especially for chirp-like signals. This diminishing return, coupled with the increased computational cost associated with more filters, led us to select $100$ filters as the current optimal setting. 

\subsection{Performance on observed GW signals}
\label{sec:results_realbbh}

\begin{figure*}[htbp]
    \centering
    \includegraphics[width=\textwidth]{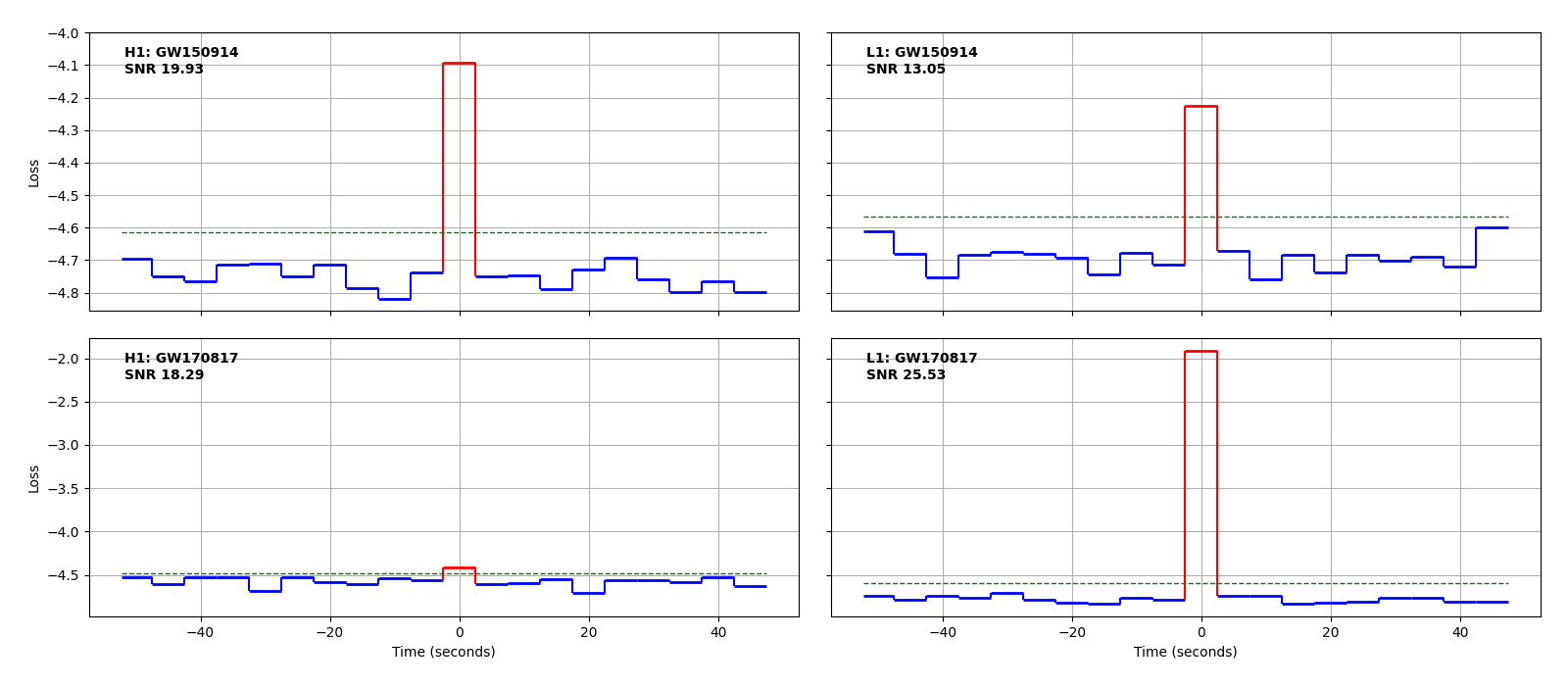}
    \caption{Loss time series for GW150914 (top) and GW170817 (bottom) in L1 (left) and H1 (right). Values exceeding the detection threshold (dashed line) are highlighted in red. The large loss spike for GW170817 in L1 corresponds to the instrumental glitch that coincided with this event. Each panel displays the single-detector SNR as reported in the 4-OGC catalog. The x-axis represents time in seconds, where $t=0$ is the reported merger time of the event.}
    \label{fig:gw_examples}
\end{figure*}

We first examine two landmark GW events in detail: GW150914 and GW170817. Figure~\ref{fig:gw_examples} shows the \algoname loss values from the short timescale analysis for these events in both H1 and L1. The analysis of GW170817 in the L1 detector shows an exceptionally strong loss spike coinciding with the known instrumental glitch that overlapped with this signal. Despite different noise conditions across detectors, \algoname identifies all four cases as anomalies, with loss values exceeding their respective detection thresholds. The weak detection of GW170817, a long duration chirp, in H1 is consistent with the results obtained from simulations in Table~\ref{table:SNRDet_table}.

\begin{figure}[htbp]
    \centering
    \includegraphics[width=\columnwidth]{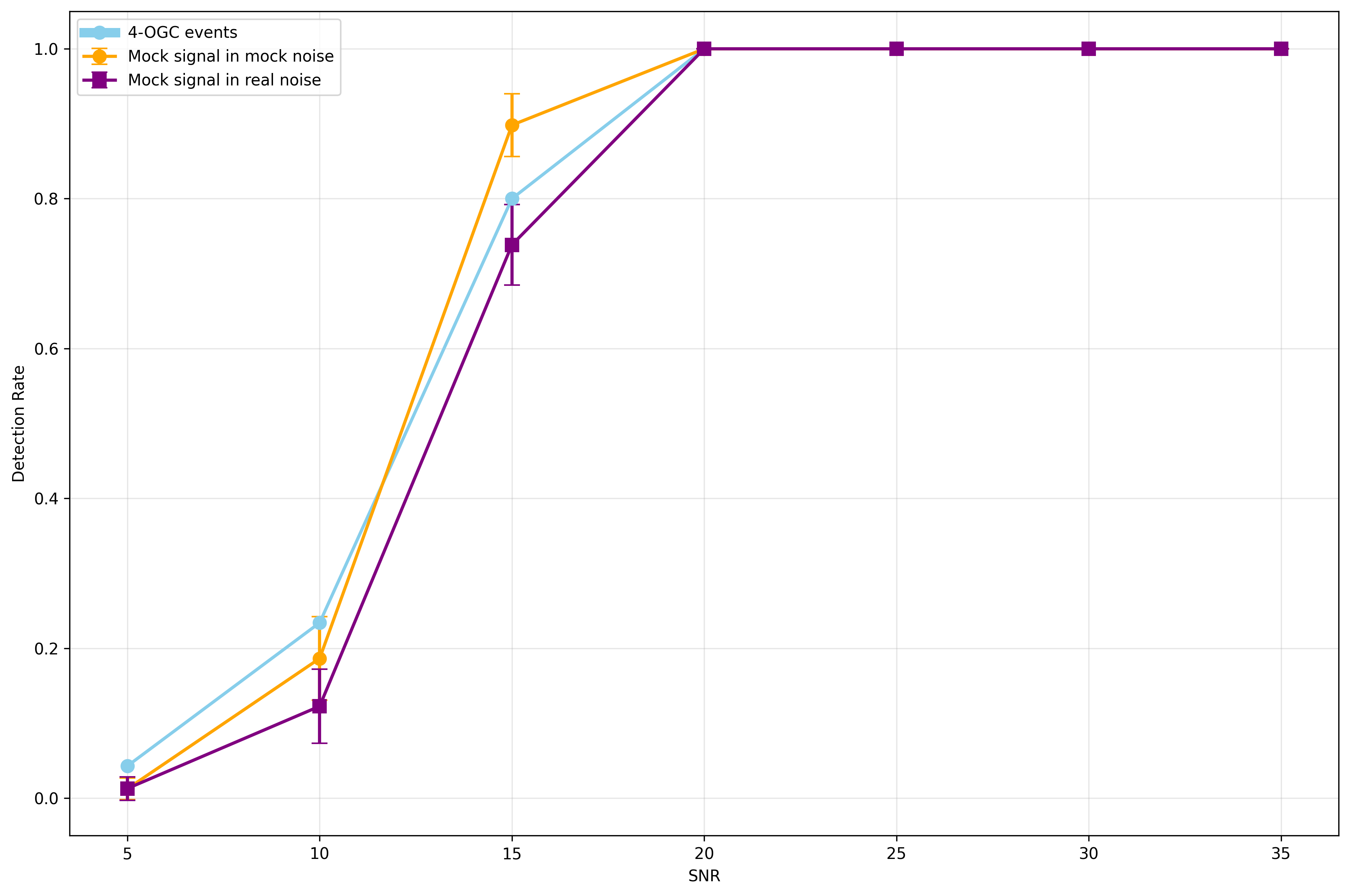}
    \caption{Detection probability versus SNR curves for three cases: mock noise with mock signals (orange), real data with mock signals (purple), and real GW events from the 4-OGC catalog binned by SNR (blue). Error bars are derived from bootstrap sampling.}
    \label{fig:snr_detection}
\end{figure}

To further evaluate the detection performance on GW anomalies, we conducted the following experiments. The objective here is to check if the detection efficiencies for real GW signals in real data are consistent with the ones measured for mock signals. For this, we use the results from the injections of mock IMRPhenomD signals in mock noise (c.f., Fig.~\ref{fig:ROC_curves}) and, to account for variations in detection efficiency caused by the properties of real data, also inject mock signals in real data.  Our evaluation focuses on events from the 4-OGC catalog~\cite{nitz20234}, specifically on signals with durations $< 5$~sec to match our short timescale analysis window. 
\begin{enumerate}
    \item \textit{Mock GW signals in real noise}: We injected IMRPhenomD waveforms into real data segments, with SNRs based on the ASDs estimated from the lowest-loss segments for the corresponding detectors and observing runs. However, the injections were not done in the lowest-loss segments but those corresponding to the mode (highest frequency) of the loss distribution. Thus, the noise realizations used in this experiment are representative of the typical noise behavior in each run, not extremes with exceptionally low or high anomaly contamination. The injections were for a discrete set of SNR values with $200$ injections per SNR. It should be noted that the actual SNR of an injected signal for a given segment may be different from these values since its PSD need not match the reference obtained from the lowest-loss segment in its observing run.
    \item \textit{Real GW signals in real noise}: We analyze the confirmed GW detections in the complete 4-OGC catalog. To account for their random SNRs, we bin them according to their single-detector SNR values in $5$ equally spaced bins in the range $[5,40]$.  The detection probability for each bin was set as the fraction of signals in that bin that were detected by \algoname.
\end{enumerate}
Figure~\ref{fig:snr_detection} presents the detection probability as a function of SNR for the two experiments above along with the results for the case of mock IMRPhenomD signals in mock non-anomalous noise given in Sec.~\ref{sec:results_real}. The results from experiment~$1$ above show a slight degradation in performance, with an increase in the error bars from our bootstrap analysis reflecting the variability in real detector noise. The binned 4-OGC events are consistent with both of the preceding analyses and the results in Table~\ref{table:SNRDet_table}, particularly near $50\%$ detection efficiency. The consistency between mock and real event detection efficiencies indicates that the performance of \algoname on simulations translates well to real signals and data. This is particularly noteworthy given that the algorithm was trained exclusively on mock non-anomalous data. 

\subsection{Non-GW anomalies}
\label{sec:results_real}

\begin{figure*}[htbp]
    \centering
    \includegraphics[width=\textwidth]{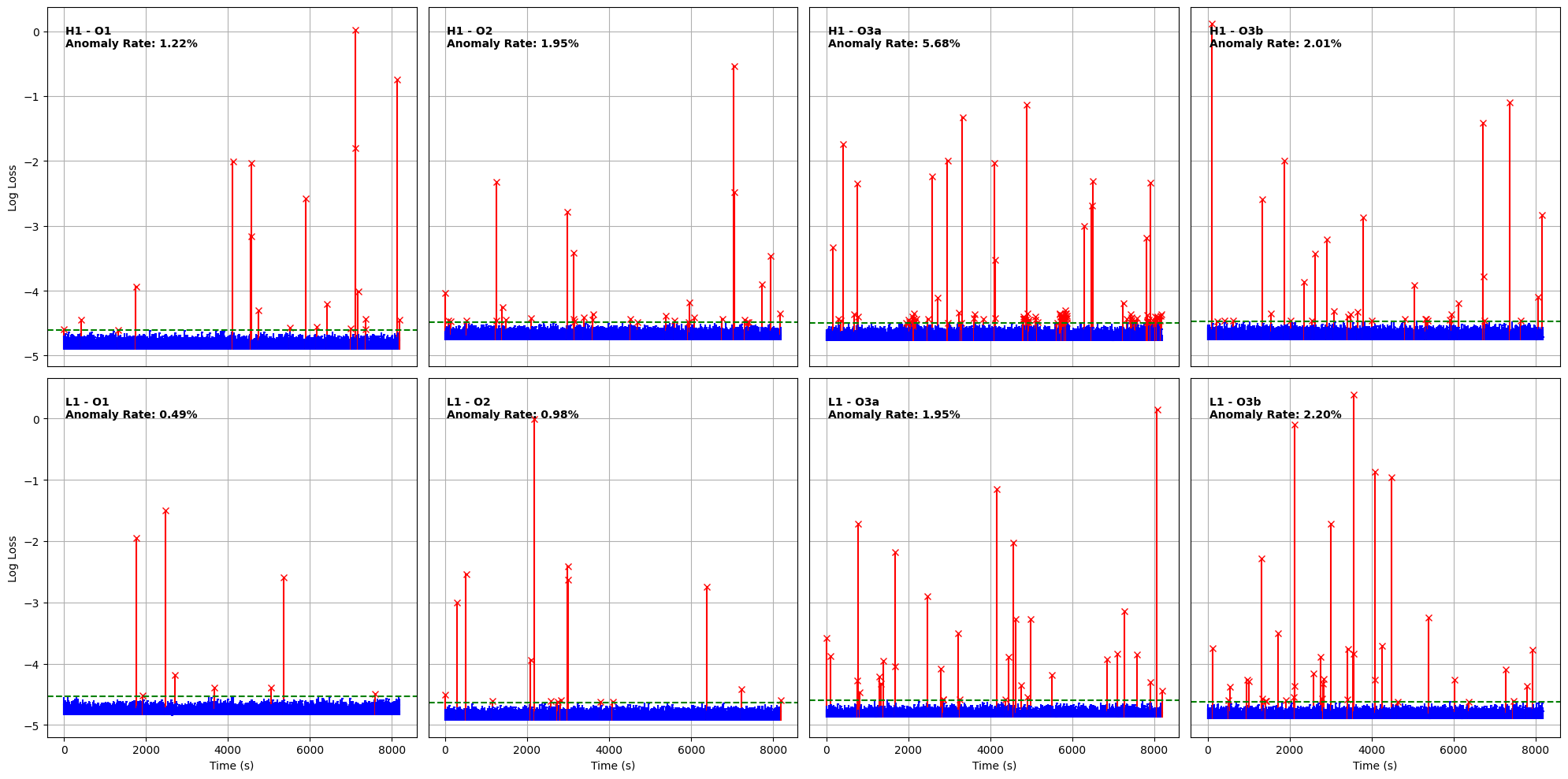}
    \caption{Time series of \algoname loss values for different detectors (rows) and observing runs (columns) using randomly chosen $8000$~sec long segments. Loss values exceeding the detection threshold (horizontal dashed line) are highlighted in red, representing detected anomalies. Given the settings used for \algoname, this  figure illustrates the temporal distribution of only short duration ($O(1)$~sec or less) anomalies.}
    \label{fig:detector_timeseries}
\end{figure*}

We used \algoname to analyze $546.00$ minutes ($9.1$ hours) of coincident data from observing runs O1, O2, O3a, and O3b, selecting time segments where both L1 and H1 were operational. In total, $270$ anomalies were detected, corresponding to an overall anomaly rate of $29.7$ per hour. This significantly exceeds the expected FAR of $\approx 3.6$~events per hour based on our detection thresholds (c.f., Sec.~\ref{sec:detection_thresholds}). However, as shown in Fig.~\ref{fig:detector_timeseries}, the rate of detected anomalies varies significantly across the observing runs, detectors, as well as in time. The rate of anomalies generally increases with the lowering of the noise Power Spectral Density (PSD) with subsequent observing runs, an observation consistent with other studies\cite{steltner2022identification}. For example, the percentage of anomalous segments in later observing runs ($1.95$\% for L1-O3a and $2.20$\% for L1-O3b) is higher than earlier runs ($0.49$\% for L1-O1 and $0.98$\% for L1-O2). This trend likely reflects the increasing sensitivity of the detectors, which uncovers more environmental and instrumental disturbances. It is also observed that H1 in O3a had an exceptionally elevated rate of anomalies; however, most of the excess can be attributed to loss values close to the detection threshold.

While Fig.~\ref{fig:detector_timeseries} illustrates the anomaly rates using representative segments, ABNORMAL was run on much longer data segments for some of the observational runs. Table~\ref{tab:anomaly_distribution} gives the complete number of anomalies found in those longer analyzed durations.
\begin{table}[htbp]
    \caption{\label{tab:anomaly_distribution} Total count of anomalous events detected across different observing runs using thresholds corresponding to a FAR of $\approx 10^{-3}$~events/sec or $3.75$~events/hr. The values in parentheses are the corresponding anomaly rates in events per hour.}
    \begin{ruledtabular}
\begin{tabular}{lccc}
Run & \makecell{Duration \\ (min)} & \makecell{H1 \\ Count (rate/hr)} & \makecell{L1 \\ Count (rate/hr)} \\
\hline
O1 & 136.50 & 20 (8.79) & 8 (3.52) \\
O2 & 136.50 & 32 (14.07) & 16 (7.03) \\
O3a & 136.50 & 93 (40.88) & 32 (14.07) \\
O3b & 136.50 & 33 (14.51) & 36 (15.82) \\
\hline
Combined & 546.00 & 178 (19.56) & 92 (10.11) \\
\end{tabular}
\end{ruledtabular}
\end{table}

\begin{figure*}[htbp]
    \centering
    \includegraphics[height=0.9\textheight]{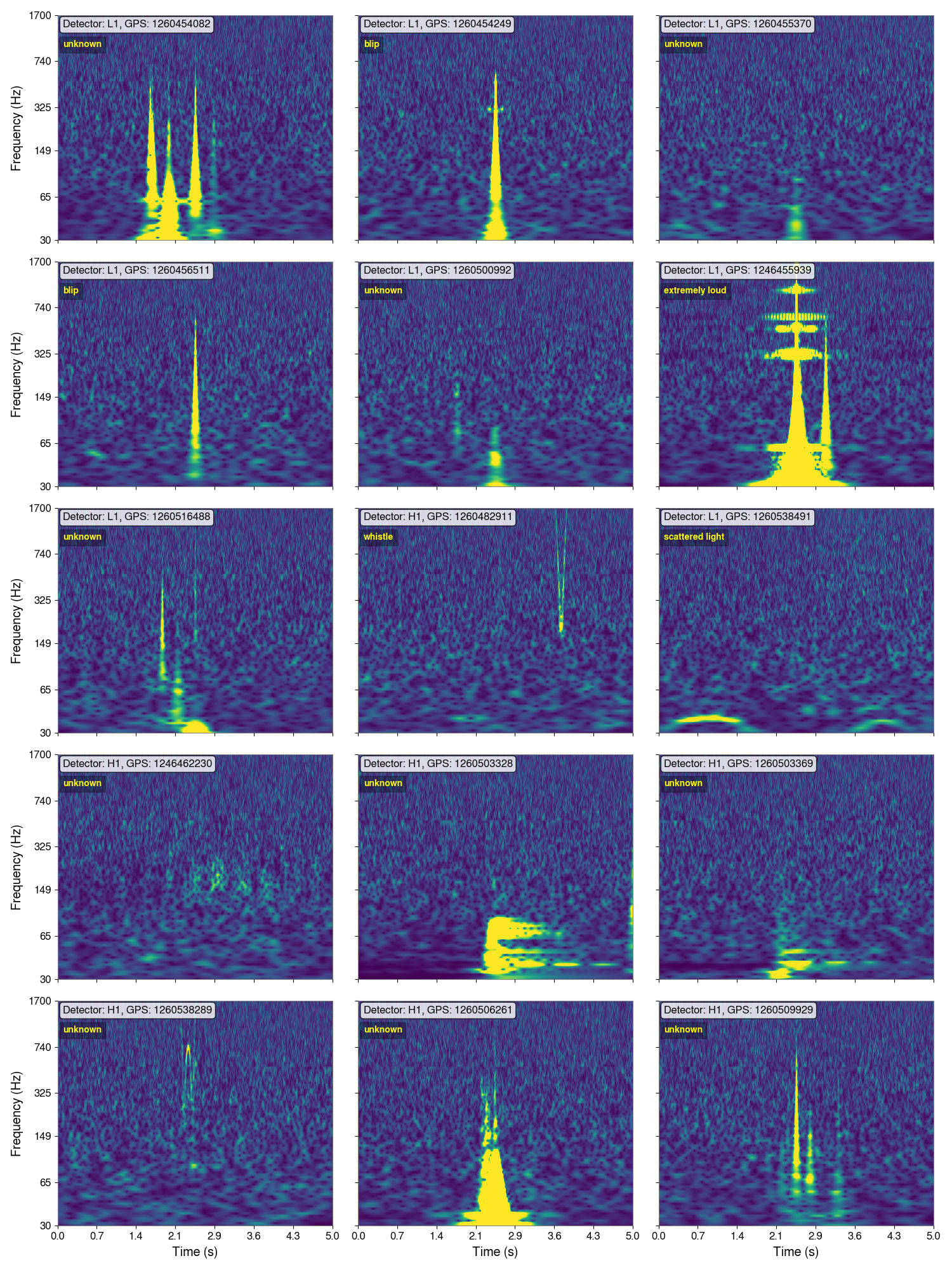}
    \caption{Examples of short timescale anomalies. Each panel shows a normalized Q-transform centered on the highest value of the time-series and contains the detector name, GPS start time, and a label for the glitch class.}
    \label{fig:anomaly_examples_real}
\end{figure*}

Figure~\ref{fig:anomaly_examples_real} presents a selection of representative non-GW anomalies detected by \algoname. The figures were generated using constant-Q transforms with Q $\approx$ 23, covering frequencies from 30 to 1700 Hz at 4096 Hz sampling rate. Each pixel at a given frequency is normalized by the median of all pixel values at that frequency.
Each example is characterized by a distinct pattern in both the time and frequency domains, demonstrating the ability of \algoname to identify a wide range of anomaly morphology. While an automated classification of these anomalies is beyond the scope of this work, visual inspection reveals glitches from known Gravity Spy classes but also many that do not. The glitches in Fig.~\ref{fig:anomaly_examples_real} are annotated with class labels according to our visual inspection, with glitches that do not belong to Gravity Spy classes labeled unknown.


\subsubsection{Long time scale non-GW anomalies}
\begin{figure}
    \centering
    \includegraphics[scale=0.3]{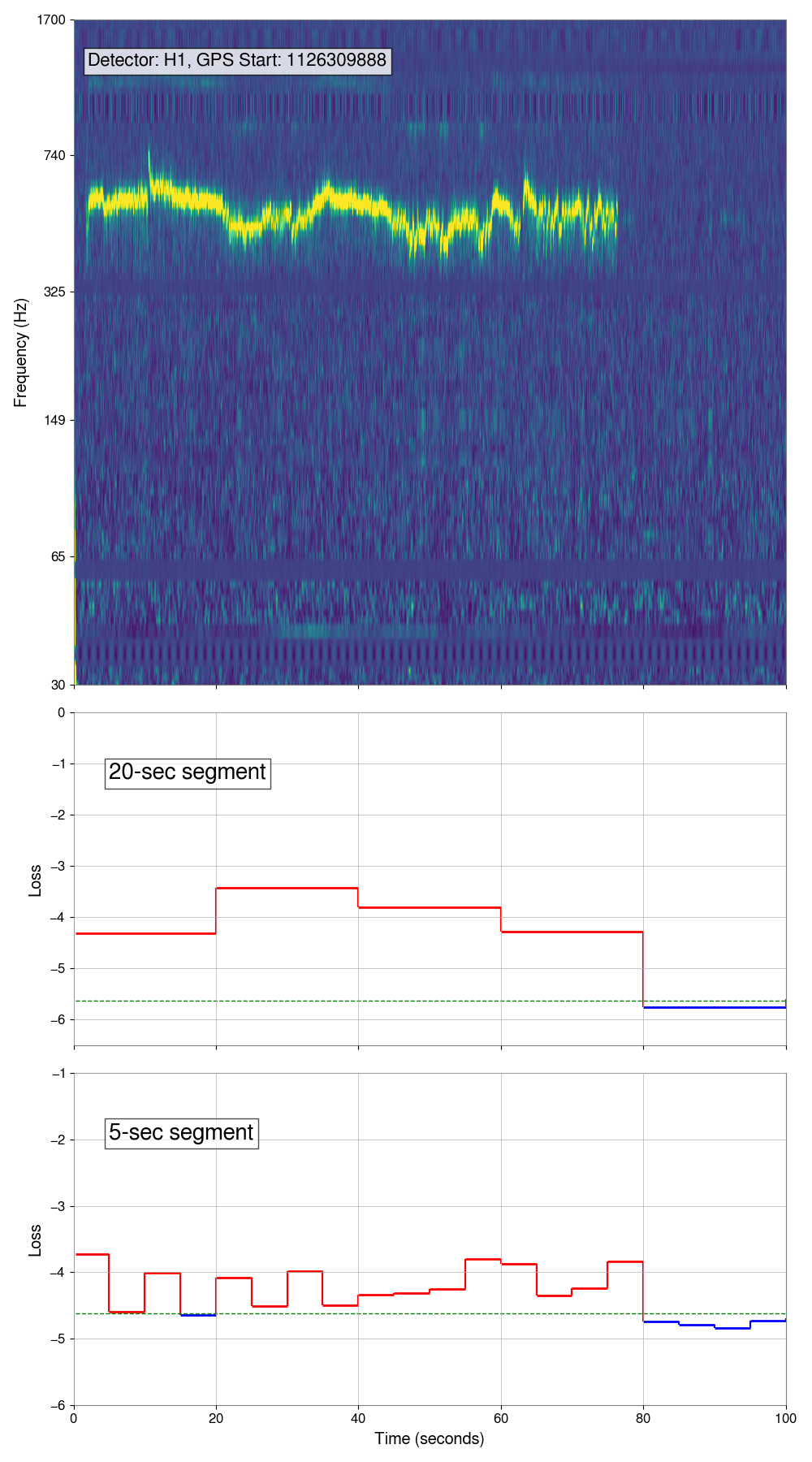}
    \caption{Detection of a wandering spectral line by ABNORMAL. The top panel shows the CQT spectrogram of the data, revealing a spectral line centered around 600~Hz that varies in frequency over time. The middle and bottom panels display the reconstruction loss values from the long and short time scale analyses. Red colored loss values exceed the detection threshold shown by the green dotted line.}
    \label{fig:wanderling-line}
\end{figure}

The long time scale analysis displays various non-GW anomalies with distinct morphologies. Figure~\ref{fig:wanderling-line} presents the analysis of one such anomaly, namely, a wandering line in more detail. In this case, both the short and long time scale analyses detect a spectral line around 600~Hz that varies in frequency over time. However, the long time scale analysis captures these features more effectively, with all loss values during the duration of the anomaly exceeding the threshold, while some values in the short time scale analysis appear very close to the threshold. 

\begin{figure*}[htbp]
    \centering
    \includegraphics[width=\linewidth]{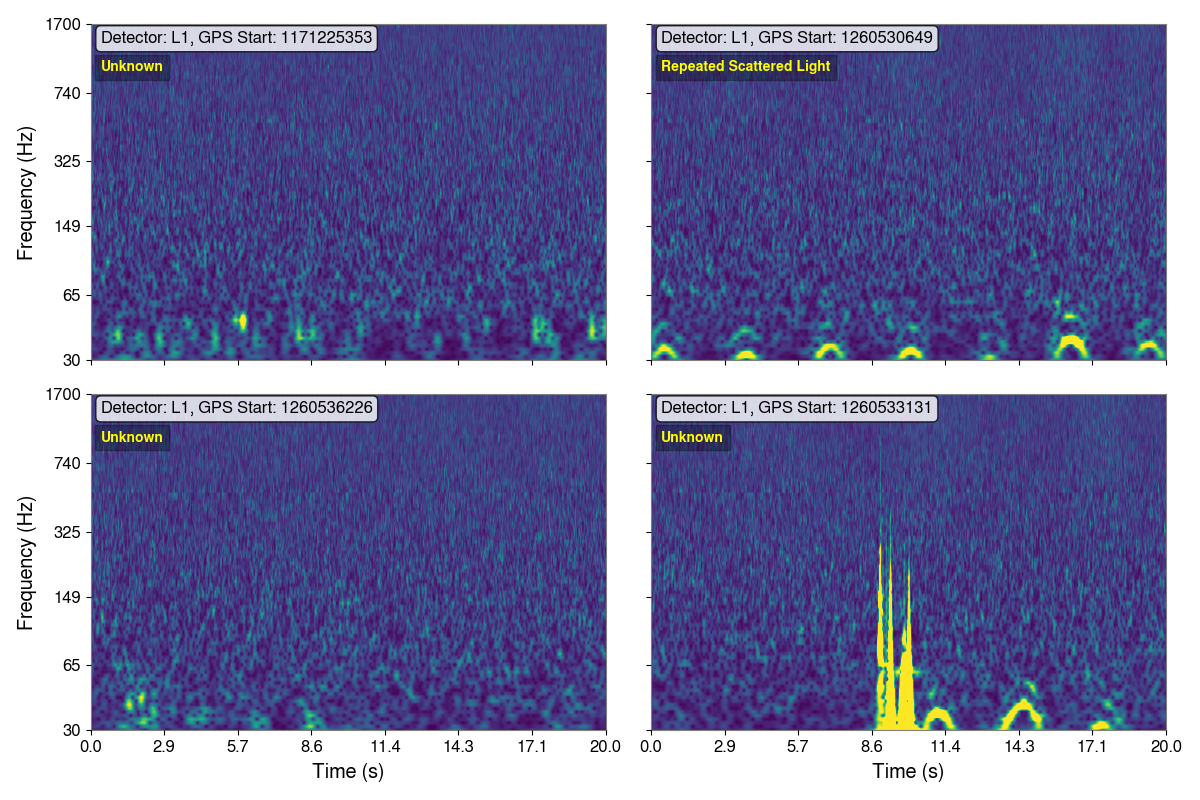} 
    \caption{Examples of long timescale anomalies. Each panel shows a normalized Q-transform centered on the highest value of the time-series and contains the detector name and GPS start time. }
    \label{fig:new_anomalies_examples} 
\end{figure*}

Further illustrating the variety of transient noise features encountered, Fig.~\ref{fig:new_anomalies_examples} presents four more examples of anomalies identified in the long time scale analysis. As in the case of short timescale anomalies, we found many that belong to existing Gravity Spy classes but also several that fall outside. 

\section{Conclusions}
\label{sec:conclusions}

We have presented \algoname, a semi-supervised anomaly detection method for GW data
that introduces the following novel features. (1) Training performed exclusively on simulated stationary noise based on design sensitivities rather than real data. This eliminates the need for data labeling, which requires ad hoc assumptions about stationary segments in real data, as well as biases in training due to undetected anomalies. (2) Use of statistical features rather than raw time series values that leverages autoencoders more effectively for pure noise data. The only data pre-processing steps required are the limiting of data to the frequencies of interest ($[30, 1700]$~Hz in this paper) and removal of strong lines using fixed notch filters.

Our results demonstrates the effectiveness of \algoname in identifying both GW signals and non-GW anomalies across different observing runs. The method successfully detects approximately 90\% of BBH mergers with single detector SNR $>14.97$ at inverse false alarm rates of $0.06$ per minute. For synthetic broadband glitches, \algoname achieves detection probabilities as shown in Table~\ref{table:SNRDet_table}.

The analysis of real LIGO data revealed a rich landscape of non-stationary behaviors, with 1482 anomalies detected across $88.74$ hours of data from O1, O2, O3a, and O3b observing runs. The anomaly rate varied significantly between observing runs, with the highest rate observed in O3a ($26.37$ per hour) and the lowest in O2 ($5.79$ per hour). The LIGO Livingston detector showed consistently higher anomaly rates in later observing runs, suggesting detector-specific characteristics of non-stationarity. These findings underscore the importance of robust anomaly detection methods for both GW searches and detector characterization efforts.

In the present version of \algoname, the necessity of line removal is the principal reason for recalibration of the threshold for anomaly detection in each observing run. A more elegant solution would be to train \algoname to recognize persistent spectral features (such as 60~Hz harmonics and violin modes) as part of the non-anomalous noise. However, each line itself is a non-stationary feature due to amplitude fluctuations and this will need to be accounted for in the features fed to \algoname. This approach, to be explored in future work, could eliminate the need for explicit line removal and potentially improve sensitivity to anomalies near line frequencies.

 Additionally, extending \algoname to process data from multiple detectors simultaneously could enhance its ability to distinguish between local instrumental artifacts and genuine astrophysical signals.
Note that GWAK uses network data, hence, it is expected to be more sensitive than our current single-detector method; GWAK also trains on data with simulated signals, making it more sensitive for specific classes of signals/glitches. In contrast, \algoname takes a more generalized approach by learning the characteristics of stationary noise without exposure to specific signal or glitch types during training. This fundamental difference in approach makes \algoname potentially more adaptable to previously unseen types of anomalies, while GWAK may have advantages in identifying known signal classes.

Future work will also explore the application of \algoname to targeted searches for specific types of GW signals, particularly those that may not be well-captured by traditional matched filtering approaches. By leveraging the sensitivity of the method to general non-stationarity, we aim to develop complementary search strategies that can identify exotic or unexpected signal morphologies in GW data.

\acknowledgements{SDM is supported by NSF Grant No.~PHY-2207935. Y-YG, XQ, and Y-XL are partially supported by the National Key Research and Development Program of China (Grants No. 2021YFC2203003 and No. 2023YFC2206701), the National Natural Science Foundation of China (Grant No. 12247101), the Fundamental Research Funds for the Central Universities (Grant No. lzujbky-2024-jdzx06), the Natural Science Foundation of Gansu Province (No. 22JR5RA389) and the ``111 Center'' under Grant No. B20063.}

\appendix
\section{Synthetic anomalies}
\label{sec:synth_anomalies}
For generating the mock anomalous test dataset, we used eight distinct classes of simulated signals. Five of the classes, labeled A to E, are constructed as superpositions of sine-Gaussian pulses following the approach in~\cite{principe2009locally, principe2017locally} and serve as models for the type of glitches, such as Blip and Tomte, identified in Gravity Spy. The waveform of a signal $G(t)$ in these classes is given by
\begin{equation}
G(t) = \sum_{i=1}^{N_p} a_i e^{-2\pi^2 f_i^2 \left(t-\Delta_i\right)^2} \cos\left[2\pi f_i \left(t-\Delta_i\right)\right]\;,
\end{equation}
where $N_p$ is the number of constituent pulses, $a_i$ is the amplitude, $f_i$ is the central frequency, and $\Delta_i$ is the time shift of the $i$-th pulse. The five classes differ in the base values of these parameters and have varying levels of complexity in their temporal and frequency characteristics. Within a class, the members have random values of $a_i$, $f_i$, and $\Delta_i$ centered around the base values and, for one class, random values of $N_p$. In all cases, the perturbations to base values are drawn independently from uniform probability density functions (pdfs). In one case, the base values themselves are drawn randomly before being perturbed. The properties of the glitch classes A to E are summarized in Table~\ref{tab:glitch_parameters}.
\begin{table*}[htbp]
    \centering
    \caption{Parameters for Synthetic Glitch Classes A through E. Here $U([a,b])$ denotes the Uniform pdf
    over $[a,b]$ if $[a,b]\subset \mathbb{R}$ or the discrete uniform probability distribution if $[a,b]\subset \mathbb{Z}^+$.}
    \label{tab:glitch_parameters}
    \begin{tabularx}{\textwidth}{|c|c|X|X|}
        \hline
        \textbf{Class} & \textbf{\makecell{$N_p$}} & \textbf{Base Parameters} & \textbf{Perturbations} \\
        \hline
        \textbf{A} & 4 & \makecell[l]{$f_i$: (400, 150, 270, 90) Hz \\ $\Delta_i$: (0.0055, -0.0065, 0, -0.017) s \\ $a_i$: (-4, 6, -2.2, 0.3)} & \makecell[l]{$\delta f_i \sim U([-10, 10])$ Hz \\ $\delta \Delta_i \sim U([-0.003, 0.003])$~s \\ $\delta a_i \sim U([-1, 1])$} \\
        \hline
        \textbf{B} & 3 & \makecell[l]{$f_i$: (150, 175, 200) Hz \\ $\Delta_i$: (0.013, -0.001, 0) s \\ $a_i$: (-1, 10, -9)} & \makecell[l]{$\delta f_i \sim U([-10, 10])$ Hz \\ $\delta \Delta_i \sim U([-0.001, 0.001])$~s \\ $\delta a_i \sim U([-1, 1])$} \\
        \hline
        \textbf{C} & 2 & \makecell[l]{$f_i$: (1050, 990) Hz \\ $\Delta_i$: (0.0005, -0.0002) s \\ $a_i$: (8, -7)} & \makecell[l]{$\delta f_1, \delta f_2 \sim U([-10, 10])$ Hz \\ $\delta \Delta_1 \sim U([-0.0003, 0.0003])$~s \\ $\delta \Delta_2 \sim U([-0.0002, 0.0002])$~s \\ $\delta a_1 \sim U([-1, 1])$ \\ $\delta a_2 \sim U([-0.2, 0.2])$} \\
        \hline
        \textbf{D} & \makecell{$U([1, 10])$} & \makecell[l]{$f_i \sim U([200, 1000])$ Hz \\ $\Delta_i \sim U([-1, 1])$~s \\ $a_i \sim U([-10, 10])$} & \makecell[l]{$\delta f_i \sim U([-10, 10])$ Hz \\ $\delta \Delta_i \sim U([-0.003, 0.003])$~s \\ $\delta a_i \sim U([-1, 1])$} \\
        \hline
        \textbf{E} & 4 & \multicolumn{2}{p{0.6\textwidth}|}{This class consists of four repeated instances of a Class A glitch. The center of this pattern has a global time shift drawn from $U([-0.5, 0.5])$~s. The four glitches within the pattern have fixed time shifts of $(-0.45, -0.15, 0.15, 0.45)$~s relative to the center. All four have common values for the perturbations, distributed as per Class A.} \\
        \hline
    \end{tabularx}
\end{table*}

The sixth and seventh classes of simulated anomalies consist of chirp signals with linear and quadratic frequency evolution, respectively. The instantaneous frequency $f(t)$ of the linear chirp signals  follows:
\begin{equation}
    f(t) = f_{\text{start}} + (f_{\text{end}} - f_{\text{start}})\frac{t}{T}\;,
    \label{eq:linear_chirp}
\end{equation}
where $f_{\rm start}$ and $f_{\rm end}$ are the initial and terminal frequencies, respectively, and $T$ represents the signal duration. Both frequencies are confined to the sensitive band of LIGO and drawn independently from uniform distributions with $f_{\text{start}} \sim U([50, 100])$~Hz and $f_{\text{end}}\sim U([500, 1200])$~Hz. For the  quadratic chirps, we maintain the same distributions while implementing a quadratic time dependence in the frequency evolution,
\begin{equation}
    f(t) = f_{\text{start}} + (f_{\text{end}} - f_{\text{start}})\left(\frac{t}{T}\right)^2\;.
    \label{eq:quadratic_chirp}
\end{equation}

\begin{figure}
    \centering
    \includegraphics[width=0.5\textwidth]{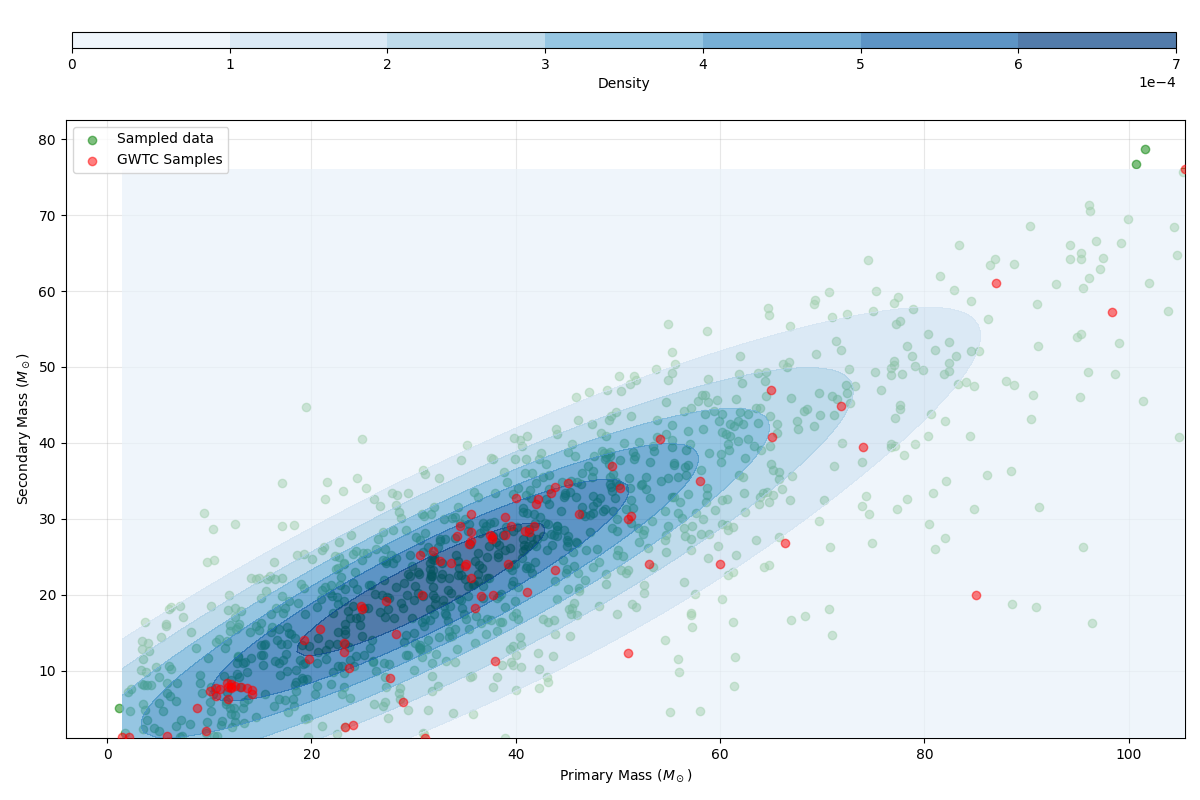}
    \caption{Comparison of the distribution of observed BBH masses (red dots) from all three GWTC catalogs and our simulated samples (green dots).
    The blue shaded regions show contours of the Kernel Density Estimate fitted to the GWTC catalogs, from which the samples are drawn using the \texttt{resample} method of a \texttt{scipy.stats.gaussian\_kde} instance.}
    \label{fig:gwtc-kde}
\end{figure}

The eighth class of signals is that of BBH mergers for which we use
the IMRPhenomD waveform model from \texttt{PyCBC}~\cite{biwer2019pycbc}. To make our simulations realistic for this case, the masses of the simulated signals are derived from the observed distribution in all three GWTC catalogs, using Kernel Density Estimation~\cite{davis2011remarks} to fit the observed distribution as shown in Fig~\ref{fig:gwtc-kde}  and sampling from the fitted distribution using the \texttt{resample} method of a \texttt{scipy.stats.gaussian\_kde} instance. Regarding the remaining signal parameters, we set the aligned spin components to $\chi_{1} = 0.8$ and $\chi_{2} = 0.4$ which are also aligned with the total angular momentum, fix the luminosity distance to $500$~Mpc, and the sky location to $\text{RA} = 1.375$~rad and $\text{DEC} = -1.2108$~rad, where RA and DEC are the right ascension and declination, respectively. To ensure computational efficiency, we retain only those generated waveforms that have durations $\leq 5$~sec when the signal is bandlimited to $[30, 1700]$~Hz.

\section{Line removal method}
\label{sec:ln_rmv_method}
We first establish a local noise floor by applying a running median filter with a block size of 1000 samples to the logarithm of the ASD. The normalized spectrum is then obtained by:
\begin{equation}
\label{eq:norm_asd}
A_{\text{norm}}(f) = \log(\text{ASD}(f)) - \text{median}(\log(\text{ASD}(f)))\;.
\end{equation}
Spectral lines are identified as peaks in this normalized spectrum that exceed a threshold of $A_{\text{norm}}(f) > 0.5$. To handle closely spaced lines, we group peaks that lie within 1~Hz of each other. For each group:
\begin{itemize}
    \item The peak with the highest magnitude is selected as the representative frequency.
    \item The bandwidth is determined by finding the frequencies where $A_{\text{norm}}(f) \geq 0.2$ on either side of the peak.
    \item Both the peak parameters (frequency and magnitude) and the bandwidth information are recorded.
\end{itemize}
Line removal is implemented using cascaded Butterworth notch filters, each having an order of $1001$. For each identified line:
\begin{itemize}
    \item The center frequency is the peak frequency of the identified band edges.
    \item The bandwidth is set to the maximum of either the identified bandwidth or 1~Hz.
    \item The pass band is set at 92\% of the bandwidth on either side of center frequency.
    \item The stop band covers the full identified bandwidth.
\end{itemize}
For lines requiring strong attenuation, we cascade multiple notch filters with moderate attenuation (controlled by a threshold of 5~dB) rather than applying a single high-attenuation filter. This approach helps maintain numerical stability and reduces ringing artifacts. Each filter is applied using forward-backward filtering (via the Python routine \texttt{scipy.signal.filtfilt}) to ensure zero-phase distortion. The cascade continues until the desired total attenuation for each line is achieved. 
To avoid edge effects from the filtering process, we crop $(N+1)/2 = 501$ samples from both ends of the filtered data, where $N=1001$ is the filter order. This ensures that transients from the filtering process do not contaminate the final output. 


\end{document}